\documentclass[fleqn,usenatbib]{mnras}

\usepackage{amsmath}
\usepackage{txfonts}                
\usepackage{graphicx}               
\usepackage[T1]{fontenc}
\usepackage{microtype}

\hypersetup{linkcolor=red}          
\newcommand{\dd}{{\,\mathop{\kern0pt\mathrm{d}}\!{}}}

\def\equationautorefname~#1\null{equation~(#1)\null}

\title[Spherically-aligned Jeans equations]{Efficient solution of the anisotropic spherically-aligned axisymmetric Jeans  equations of stellar hydrodynamics for galactic dynamics}

\author[M.~Cappellari]{Michele Cappellari\thanks{E-mail:
	michele.cappellari@physics.ox.ac.uk}\\
Sub-Department of Astrophysics, Department of Physics, University of Oxford, Denys Wilkinson Building, Keble Road, Oxford, OX1 3RH, UK}

\date{Accepted 2020 April 2. Received 2020 April 1; in original form 2019 July 22}

\pagerange{\pageref{firstpage}--\pageref{lastpage}} \pubyear{2020}

\begin{document}
\label{firstpage}
\maketitle

\begin{abstract}
    I present a flexible solution for the axisymmetric Jeans equations of stellar hydrodynamics under the assumption of an anisotropic (three-integral) velocity ellipsoid aligned with the spherical polar coordinate system. I describe and test a robust and efficient algorithm for its numerical computation. I outline the evaluation of the intrinsic velocity moments and the projection of all first and second velocity moments, including both the line-of-sight velocities and the proper motions. This spherically-aligned Jeans Anisotropic Modelling (JAM$_{\rm sph}$) method can describe in detail the photometry and kinematics of real galaxies. It allows for a spatially-varying anisotropy, or stellar mass-to-light ratios gradients, as well as for the inclusion of general dark matter distributions and supermassive black holes. The JAM$_{\rm sph}$ method complements my previously derived cylindrically-aligned JAM$_{\rm cyl}$ and spherical Jeans solutions, which I also summarize in this paper. Comparisons between results obtained with either JAM$_{\rm sph}$ or JAM$_{\rm cyl}$ can be used to asses the robustness of inferred dynamical quantities. As an illustration, I modelled the ATLAS$^{\rm 3D}$ sample of 260 early-type galaxies with high-quality integral-field spectroscopy, using both methods. I found that they provide statistically indistinguishable total-density logarithmic slopes. This may explain the previously-reported success of the JAM method in recovering density profiles of real or simulated galaxies. A reference software implementation of JAM$_{\rm sph}$ is included in the publicly-available JAM software package.
\end{abstract}


\section{Introduction}
\label{sec:intro}

\subsection{Dynamical modelling methods}

We live in a very interesting Universe. According to our current understanding, some of its key constituents do not directly emit electromagnetic radiation. For this reason, their masses or distribution can only be quantified through gravitational interactions or equivalently, by their curvature of space-time. One dark component is the mysterious dark matter, which, despite being a key piece of our model of the Universe \citep[e.g.][]{Blumenthal1984}, has been recently experiencing an existential `crisis' due to the lack of viable candidate particles, despite enormous efforts to look for them \citep[see review by][]{Bertone2018}. The other dark components are supermassive black holes in galaxy nuclei. For them, strong evidence does exist, and in the past few decades, they have been promoted from mere physical curiosity to a key element in galaxy evolution \citep[see review by][]{Kormendy2013review}. Additional nearly-dark components are stellar remnants (stellar black holes and neutron stars) and low mass stars, whose fractional contributions depends on the stellar Initial Mass Function, (IMF) which seems to be varying among different galaxies \cite[e.g.][]{vanDokkum2010,Cappellari2012} and affects our understanding of galaxy evolution. The dark components are best studied using either galaxy dynamics \citep[e.g.][hereafter BT]{Binney1987} or gravitational lensing \citep[see review by][]{Treu2010review}. This paper deals with the former technique.

Earlier dynamical models \citep[e.g.][]{Satoh1980,Binney1990,vanderMarel1990,Emsellem1994} assumed axisymmetry and were based, due to their simplicity and computational efficiency, on the equations that \citet{Jeans1922} described as ``hydrodynamical equations of motion for the stars''. These initial models additionally relied on the assumption of a semi-isotropic velocity ellipsoid ($\sigma_R=\sigma_z$ and $\overline{v_R v_z}=0$), which is a characteristic of models where the distribution function (DF) only depends on the two classic isolating integral of motion. The knowledge that the DF of galaxies depends on three integrals \citep{Ollongren1962,Contopoulos1963}, combined with the empirical finding that indeed $\sigma_R\neq\sigma_z$ in a large sample of real galaxies \citep{vanDerMarel1991}, motivated the development of the more general \citet{Schwarzschild1979} orbit-superposition dynamical models \citep[e.g.][]{Richstone1988,vanderMarel1998,Gebhardt2000,Cappellari2006,vanDenBosch2008}, including the related ``torus mapper'' technique \citep{Binney2016} and the \citet{Syer1996} ``made-to-measure'' particle-based models \citep[e.g.][]{deLorenzi2007,Dehnen2009,Long2010}.

The first and major fundamental problem when modelling external galaxies is the non-uniqueness of the surface brightness deprojection, which affects any technique \citep{Rybicki1987}. It is already severe in the axisymmetric limit at a low inclination (e.g. \citealt{Lablanche2012} and \autoref{sec:deprojection}) and becomes even more important from any viewing direction in triaxiality \citep{Gerhard1996triax}. A second problem is the fact that the observations can provide at best a three-dimensional data-cube, when using state-of-the-art integral-field stellar kinematics \citep[see review by][]{Cappellari2016}, and, for dimensional arguments alone, this cannot be expected to tightly constrain both the three-dimensional DF and the gravitational potential or galaxy shapes \citep[e.g. sec.~3 of][]{Valluri2004}. A third issue, which is often ignored, is that dynamical modelling methods only represent an approximate and, in the case of orbit or particle-based methods, a severely-discretized solution of the original mathematical problem.

Even in an ideal situation, with noiseless integral-field data, where one artificially removes the mass deprojection non-uniqueness and assumes the intrinsic mass is perfectly known, numerical experiments have revealed that one still cannot robustly recover a basic parameter like the galaxy inclination \citep{Krajnovic2005,vandenBosch2009}. Similar results were found when modelling real galaxies \citep{,Cappellari2006,deLorenzi2009}.

The severity of these degeneracies, supported by additional extensive experiments with Schwarzschild's modelling at that time, motivated my search for simpler, less-general, but hopefully more robust models, based on the Jeans equations, but this time allowing for an anisotropic (three-integral DF) $\sigma_R\neq\sigma_\phi\neq\sigma_z$  velocity ellipsoid. In \citet{Cappellari2008} I presented a very efficient Jeans solution based on the assumption of an alignment of the velocity ellipsoid in cylindrical polar coordinates. The latter approximate assumption aimed at capturing the main global characteristics of the velocity ellipsoid inferred from Schwarzschild's modelling of integral-field stellar kinematics \citep{Cappellari2007}. I dubbed the resulting method the cylindrically-aligned Jeans Anisotropic Modelling method (JAM$_{\rm cyl}$).

\subsection{Motivation for this work}
\label{sec:motivation}

On purely theoretical grounds, because of its generality, one may have expected Schwarzschild's method to be able to recover mass densities more accurately than JAM$_{\rm cyl}$. However, recent studies suggest that the reverse is true in practice, using both real galaxies and N-body simulations.

The first study used 54 real early-type and spiral galaxies for which the {\em true} circular velocity $v_c$ was assumed to be traced by the gas rotation velocity measured from the CO emission lines by the EDGE-CALIFA survey \citep{Bolatto2017}. These $v_c$ were compared against those independently obtained by fitting either Schwarzschild's or the JAM$_{\rm cyl}$ dynamical models to the same CALIFA \citep{Sanchez2012} stellar kinematics. The study found that the $v_c$ inferred using the JAM$_{\rm cyl}$ method agree more closely with the true $v_c$, than those inferred using Schwarzschild's method, especially at large radii where the gas velocities are better-determined \citep[fig.~8 of][]{Leung2018}.

The second work used N-body simulations. A direct comparison between JAM$_{\rm cyl}$ and Schwarzschild's methods was performed by \citet{Jin2019} using the currently state-of-the-art Illustris cosmological N-body simulation  \citep{Vogelsberger2014}. In this case, the {\em true} density profiles are known, as they can be inferred directly from the N-body particles. Consistently with the study on real galaxies, also this work found that the total enclosed masses $M_{\rm tot}(R)$ recovered by JAM$_{\rm cyl}$ agree more accurately with the true $M_{\rm tot}(R)$, than those inferred using Schwarzschild's method, on the same set of simulated galaxies and for the same set of adopted viewing directions \citep[fig.~6 of][]{Jin2019}. 

Of course, masses and density profiles are not the only useful metric to test and compare dynamical modelling methods. As an example, Schwarzschild's method non-parametric description of the DF can become crucial, with very high-quality data and especially for nearly edge-on galaxies, when one is trying to explicitly decompose galaxies into stellar orbital families according to their integrals of motions \citep[e.g.][]{ZhuLing2018} or stellar population \citep[e.g.][]{Long2018,Poci2019}. I do not intend to review all characteristics of the different modelling methods here.

The above reliability tests demonstrate the usefulness of the JAM technique and its complementarity to Schwarzschild's approach, even where more general methods are available and computationally feasible. These results motivate further developments in Jeans's approach which are the focus of this paper. Moreover, the availability of different Jeans methods allows for crucial tests of the sensitivity of the results to the modelling assumptions.

More specifically, the impetus for the present work comes from the existence of the Gaia DR2 data \citep{Gaia2018_DR2}, which provide three-dimensional positions and velocities for millions of stars in our Milky Way galaxy. At a significant height above the Galaxy equatorial plane, one expects the cylindrical-alignment assumption to become inaccurate as discussed in \autoref{sec:align}. This theoretical expectation was confirmed by recent Gaia studies which found that the velocity ellipsoid is well approximated by an alignment with the spherical polar coordinate system, both in the outer stellar halo \citep{Wegg2019} and in the disk region \citep{Hagen2019,Everall2019}. These data motivates the development of a practically-usable spherically-aligned solution for the Jeans equations, which we already successfully applied to the Gaia data \citep*{Nitschai2020}.

\section{General Jeans solution}

\subsection{The collisionless Boltzmann equation}

The positions $\mathbf{x}$ and velocities $\mathbf{v}$ of a large system of stars can be described by the distribution function (DF) $f(\mathbf{x},\mathbf{v})$. When the system has reached near equilibrium and is in a steady state under the gravitational influence of a smooth potential $\Phi$, the DF must satisfy the fundamental equation of stellar dynamics, the steady-state collisionless Boltzmann equation (BT equation 4-13b)
\begin{equation}\label{eq:boltzmann}
\sum _{i=1}^3 \left(v_i
\frac{\partial f}{\partial
	x_i}-\frac{\partial \Phi
}{\partial x_i}
\frac{\partial f}{\partial
	v_i}\right)=0.
\end{equation}
Given that $f$ is a function of six variables, \autoref{eq:boltzmann} is satisfied by an infinite family of solutions. One needs additional assumptions and simplifications for a practical application of the equation. One classic way of constraining the problem consists of drastically reducing it, from that of recovering the DF to that of studying only the velocity moments of the DF. This approach leads to the Jeans equations, which are discussed in the next section.

\subsection{The Jeans equations in spherical coordinates}
\label{sec:jeans_sph}

\begin{figure}
	\includegraphics[width=\columnwidth]{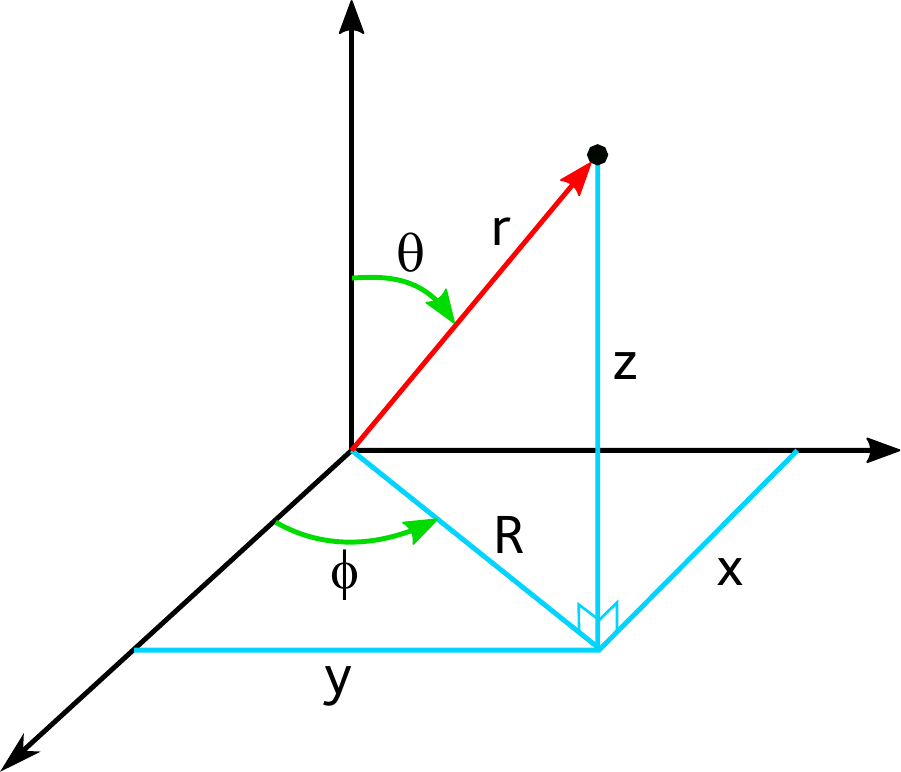}
	\caption{Definition of the spherical polar $(r,\theta,\phi)$, cylindrical polar $(R,\phi,z)$ and Cartesian $(x,y,z)$ coordinate systems adopted in this paper\label{fig:Spherical_Coordinates}.}
\end{figure}

By rewriting \autoref{eq:boltzmann} in standard spherical polar coordinates  $(r,\theta,\phi)$ (\autoref{fig:Spherical_Coordinates}) and making the important assumption of {\em axial symmetry} ($\partial \Phi/\partial\phi=\partial f/\partial\phi=0$), with $\theta=0$ on the axis of symmetry, one obtains (e.g. BT problem 4-3)
\begin{multline}\label{eq:boltz_sph}
0 = v_r \frac{\partial f}{\partial r} + \frac{v_\theta}{r} \frac{\partial f}{\partial \theta}
+ \left(\frac{v^2_\theta+v^2_\phi}{r} - \frac{\partial \Phi}{\partial r}\right) \frac{\partial f}{\partial v_r}\\
+\frac{1}{r}\left(\frac{v^2_\phi}{\tan\theta} - v_rv_\theta - \frac{\partial \Phi}{\partial \theta}\right)\frac{\partial f}{\partial v_\theta} - \frac{v_\phi}{r}\left(v_r + \frac{v_\theta}{\tan\theta} \right) \frac{\partial f}{\partial v_\phi}
\end{multline}
Multiplication of \autoref{eq:boltz_sph} respectively by $v_r$ and by $v_\theta$, and integration over all velocities, gives the two\footnote{The third Jeans equation, involving a multiplication by $v_\phi$, is not useful.}  \citet{Jeans1922} equations in spherical coordinates \citep[e.g.][equation~2.4]{deZeeuw1996}
\begin{subequations}\label{eq:jeans_sph}
	\begin{align}
	\frac{\partial(\nu\overline{v_r^2})}{\partial r} + \frac{1}{r}\left[\frac{\partial(\nu\overline{v_r v_\theta})}{\partial \theta}
	+ 2\nu\overline{v_r^2} - \nu\overline{v_\theta^2} - \nu\overline{v_\phi^2} 
	+ \frac{\nu\overline{v_r v_\theta}}{\tan\theta} \right] & = -\nu\frac{\partial \Phi}{\partial r}\label{eq:jeans_sph_r}\\
	r\,\frac{\partial(\nu\overline{v_r v_\theta})}{\partial r} + \frac{\partial(\nu\overline{v_\theta^2})}{\partial \theta}
	+ 3\nu\overline{v_r v_\theta} + \frac{\nu\overline{v_\theta^2} - \nu\overline{v_\phi^2}}{\tan\theta} & = -\nu\frac{\partial \Phi}{\partial \theta}\label{eq:jeans_sph_th}
	\end{align}
\end{subequations}
where I use the notation
\begin{equation}
\nu\overline{v_k v_j}\equiv\int v_k v_j f\; \dd^3 \mathbf{v}.
\end{equation}
\citet{Wegg2019} used \autoref{eq:jeans_sph} to infer the gravitational force field of the Milky Way using Gaia DR2 data and concluded that the gravitational potential of the dark matter is nearly spherical.

These equations are still quite general, as they derive from the steady-state Boltzmann \autoref{eq:boltzmann} with the only assumption of axisymmetry. They do {\em not} require self-consistency (a potential $\Phi$ generated by the luminosity density $\nu$) and they make no assumptions on the DF. However, even if one assumes $\Phi$ to be known (it may be derived from the observed $\nu$ via the Poisson equation), the two \autoref{eq:jeans_sph} are still a function of the four unknown $\overline{v_r^2}$, $\overline{v_\theta^2}$, $\overline{v_\phi^2}$ and $\overline{v_r v_\theta}$ and do not uniquely specify a solution.

\subsection{On the alignment of the velocity ellipsoid}
\label{sec:align}

To obtain a unique solution for the axisymmetric Jeans equations one needs to assume a shape and orientation for the velocity ellipsoid. In \citet{Cappellari2008} I reviewed the possible natural choices for the alignment of the velocity ellipsoid, namely (i) prolate spheroidal coordinates, (ii) spherical coordinates and (iii) cylindrical ones. I pointed out that real galaxies {\em cannot} be described globally neither by spherically-aligned nor by cylindrical-aligned solutions. Instead, the velocity ellipsoid must be aligned in a coordinate system qualitatively similar to the prolate-spheroidal one \citep[fig.~1 of][]{Cappellari2008}. 

The alignment of the velocity ellipsoid, unlike its axial ratios, is a characteristic of the gravitational potential alone. It contains no information on the dynamical status of the galaxy or its past evolution. In fact, for an assumed axisymmetric gravitational potential, a description of the alignment of the velocity ellipsoid can be determined numerically without a dynamical model by simply integrating orbits in that potential. The velocity ellipsoid must be aligned with the envelopes of the orbits in the $(R,z)$ meridional plane \citep[e.g. fig.~6 of][]{Cappellari2006} because along the principal axes of the velocity ellipsoid it must be possible, for the regular orbits, to approximate the orbital motions as a linear combination of two independent oscillations (plus a rotation around $\phi$) \citep{Eddington1915}. 

The orbital envelopes are radially oriented only when the potential is spherical, as in that case, the orbits are planar. The envelopes are cylindrically oriented only when the potential is plane-parallel, as in that case, the amplitude of the `vertical' $z$ oscillation is independent of cylindrical radius $R$. This implies that a spherical alignment of the velocity ellipsoid is only possible for spherical potentials and a cylindrical alignment for plane-parallel ones. These expectations were proven analytically by \citet{Evans2016}, who also showed that alignment in strictly prolate-spheroidal coordinates only holds for separable or St\"ackel potentials.

Given that no real galaxy is either a sphere, a plane parallel distribution, or has a separable potential, does this imply any of those assumptions is unphysical and not useful for the dynamical modelling of real galaxies? The answer to this question must rely on actual measurements rather than purely theoretical arguments. After all, science invariably relies on sensible approximations of reality. No real galaxy is in a steady-state, nor has a simple spherical, axisymmetric or triaxial shape as the dynamical models invariably assume. Nonetheless, approximated dynamical modelling proved very useful: They allowed us to learn e.g. about supermassive black holes \citep[see review by][]{Kormendy2013review}, dark matter \citep[see review by][]{Courteau2014} and orbital distributions \citep[see review by][]{Cappellari2016} in galaxies. The usefulness of a dynamical modelling approach must be quantified by its ability to measure the physical quantities one is interested in studying as discussed in \autoref{sec:intro}.

\subsection{Spherically-aligned Jeans solution}
\label{sec:spherical_jeans_solution}

To find a solution for the Jeans equations I start from \autoref{eq:jeans_sph} and assume that the velocity ellipsoid is aligned with the spherical coordinate system. The cross-terms of the second velocity moment tensor vanish and the Jeans equations become 
\begin{subequations}\label{eq:jeans_sph_align}
	\begin{align}
	\frac{\partial(\nu\overline{v_r^2})}{\partial r} + \frac{2\nu\overline{v_r^2} - \nu\overline{v_\theta^2} - \nu\overline{v_\phi^2}}{r} & = -\nu\frac{\partial \Phi}{\partial r}\label{eq:jeans_sph_align_r}\\
	\frac{\partial(\nu\overline{v_\theta^2})}{\partial \theta} + \frac{\nu\overline{v_\theta^2} - \nu\overline{v_\phi^2}}{\tan\theta} & = -\nu\frac{\partial \Phi}{\partial \theta}\label{eq:jeans_sph_align_th}.
	\end{align}
\end{subequations}
\citet{Bowden2016} pointed out that \autoref{eq:jeans_sph_align_th} ``does not involve the radial velocity dispersion at all'' and solved it by itself to study the flattening of the gravitational potential. Their solution involves expanding in a Fourier series the angular variation of the $\overline{v_\phi^2}/\overline{v_\theta^2}$ ratio. A feature of this approach is that one needs to specify a boundary condition in $\overline{v_\theta^2}$ (they obtain this from the data) at the adopted radius rather than specifying the usual boundary condition at infinity.

Here I follow the more common approach and look for a global solution. For this, I define the anisotropy as
\begin{equation}
\beta = 1 - \overline{v_\theta^2}/\overline{v_r^2}= 1 - \sigma_\theta^2/\sigma_r^2\\
\end{equation}
the Jeans \autoref{eq:jeans_sph_align} become \citep[e.g.][eq.~1, 2]{Bacon1983}
\begin{subequations}
	\begin{align}
	\frac{\partial(\nu\overline{v_r^2})}{\partial r} + \frac{(1+\beta)\,\nu\overline{v_r^2} - \nu\overline{v_\phi^2}}{r} & = -\nu\frac{\partial \Phi}{\partial r}\label{eq:jeans_beta_r}\\
	(1-\beta)\frac{\partial(\nu\overline{v_r^2})}{\partial \theta}  
	+ \frac{(1-\beta)\,\nu\overline{v_r^2} - \nu\overline{v_\phi^2}}{\tan\theta} & = -\nu\frac{\partial \Phi}{\partial \theta}\label{eq:jeans_beta_th}.
	\end{align}
\end{subequations}
I eliminate $\nu\overline{v_\phi^2}$ between the two equations, obtaining
\begin{equation}\label{eq:differential_r}
\frac{(1-\beta)\tan\theta}{r}\, \frac{\partial(\nu\overline{v_r^2})}{\partial \theta}
- \frac{2\beta\,\nu\overline{v_r^2}}{r}
- \frac{\partial(\nu\overline{v_r^2})}{\partial r} = \Psi(r,\theta)
\end{equation}
where I defined
\begin{equation}\label{eq:Phi}
\Psi(r,\theta) = \nu(r,\theta)\times\left(\frac{\partial\Phi}{\partial r} 
- \frac{\tan\theta}{r}\frac{\partial\Phi}{\partial\theta}\right).
\end{equation}
Now \autoref{eq:differential_r} is a linear first-order partial differential equation for $\nu\overline{v_r^2}(r,\theta)$ in two independent variables for which well-established procedures of solution exist. It can be solved with the method of characteristics \citep[e.g. section~9.2 of][]{Arfken2013mathematical} and a detailed solution was given by \citet{Bacon1983} and \citet{Bacon1985}. I now make the key assumption that the anisotropy $\beta$ is spatially constant\footnote{As will become clear later, the constant anisotropy assumption only applies to an individual component of my expansion, not to the whole galaxy. The final solution will allow for general spatial variations of the anisotropy.}. Moreover I assume the natural boundary condition that $\nu\overline{v_r^2}=0$ as $r\rightarrow\infty$. Note that this condition is much less restrictive than requiring $\overline{v_r^2}=0$ as $r\rightarrow\infty$ because the tracer density $\nu$ decreases much faster than the velocity dispersion in real galaxies. Written explicitly, the solution reads
\begin{subequations}\label{eq:sig_r}
	\begin{align}
	&\nu\overline{v_r^2}(r,\theta) = \int_r^\infty \left(\frac{r'}{r}\right)^{2\beta} \Psi(r',\theta')\; \dd r'\\
	&\theta'=\arcsin\left[\left(\frac{r'}{r}\right)^{\beta-1}\!\!\sin\theta\right].
	\end{align}
\end{subequations}
After obtaining $\nu\overline{v_r^2}$, the second moment in the tangential direction is derived e.g. from \autoref{eq:jeans_beta_th} as
\begin{equation}\label{eq:v2_phi}
\nu\overline{v_\phi^2}(r,\theta) = (1-\beta)\left[\nu\overline{v_r^2} + \frac{\partial(\nu\overline{v_r^2})}{\partial \theta}\tan\theta\right]
+ \nu\,\frac{\partial \Phi}{\partial \theta}\tan\theta
\end{equation}
By definition the other components of the second velocity moment tensor, and the mean velocity, are given by
\begin{subequations}
	\begin{align}
	\overline{v_\theta^2} &= (1 - \beta)\,\overline{v_r^2}\\
	\sigma_\phi^2 &= (1 - \gamma)\,\overline{v_r^2}\\
	\overline{v_\phi}^2 &= \overline{v_\phi^2} - \sigma_\phi^2\label{eq:v2phi_split}
	\end{align}
\end{subequations}

\begin{figure}
	\includegraphics[width=\columnwidth]{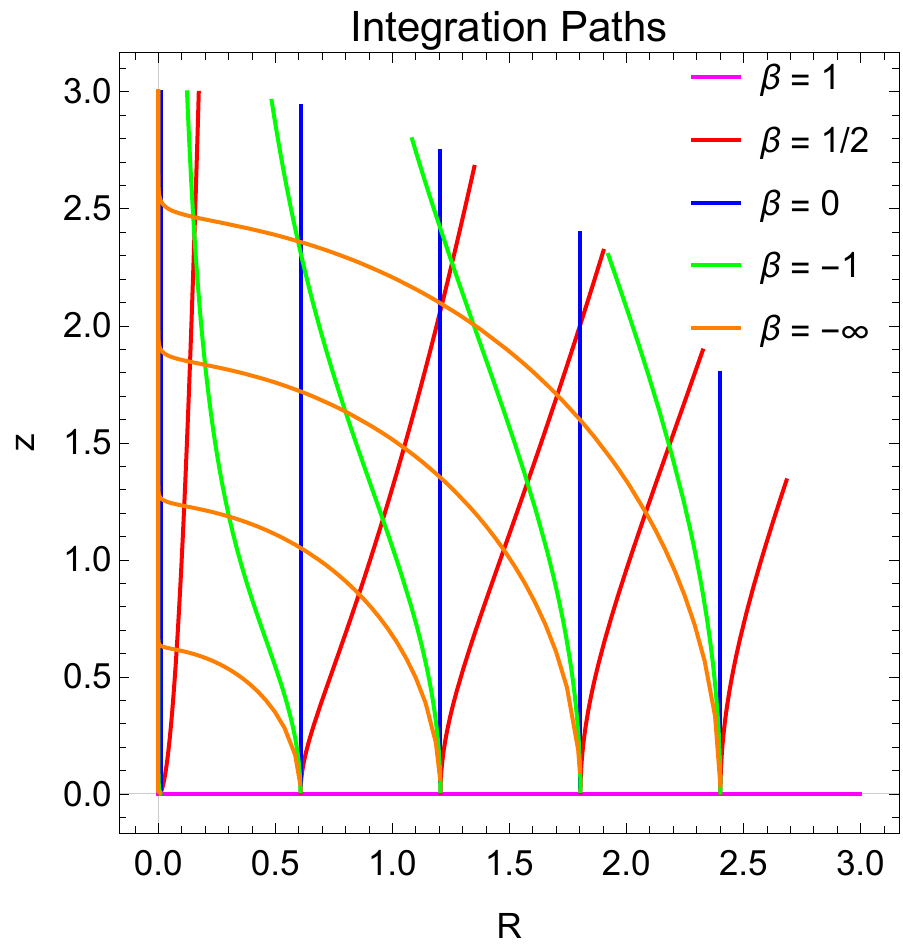}
	\caption{Integration paths for the Jeans solution, for points at different radii $R$ along a galaxy equatorial plane $z=0$. The value of the solution at a given $R$ is uniquely determined by the values of the tracer density and the gravitational potential along that curve. Different colours refer to different anisotropies, as given in the figure legend. \label{fig:integration_path}}
\end{figure}

In the spherical limit $\partial\Phi/\partial\theta=0$ and \autoref{eq:sig_r} reduces, as expected, to the spherical solution of \autoref{eq:spherical_jeans}
\begin{equation}
\nu\overline{v_r^2}(r) = \int_r^\infty \left(\frac{r'}{r}\right)^{2\beta} \nu(r')\frac{\dd \Phi(r')}{\dd r'}\; \dd r'
\end{equation}
while in the general axisymmetric case, on the symmetry $z$-axis, $\tan\theta=0$ and the solution becomes
\begin{equation}
\nu\overline{v_r^2}(r,0) = \int_r^\infty \left(\frac{r'}{r}\right)^{2\beta} \nu(r',0)\frac{\partial\Phi(r',0)}{\partial r'}\; \dd r',
\end{equation}
which is the same solution as for a spherical anisotropic model that has the same $\Phi(r')=\Phi(r',0)$ and $\nu(r')=\nu(r',0)$ radial profile as the axisymmetric model along the symmetry axis. This is useful for testing and to get a qualitative sense of the solutions.
In the semi-isotropic limit $\beta=0$ the solution reduces to the cylindrically-aligned one of \autoref{eq:jeans_sol_z}
\begin{equation}
\nu\overline{v_r^2}(R,z) = \int_z^\infty \nu(R,z)\frac{\partial\Phi(R,z)}{\partial z}\; \dd z,
\end{equation}

To interpret a dynamical model it is instructive to consider the integration path of \autoref{eq:sig_r}, in the galaxy meridional plane. To compute the solution at a given position, the galaxy density and gravitational potential are only sampled along this curve and no information on the density and potential can be inferred outside of this path. The integration curves for points along the galaxy equatorial plane, for different anisotropies, are shown in \autoref{fig:integration_path}. As one may have expected, the path is radially oriented in the limit of purely radial orbits $\beta=1$, it is parallel to the symmetry $z$-axis, for semi-isotropy $\beta=0$ as in the cylindrically-aligned solution, and is along circles for purely tangential orbits $\beta=-\infty$  (and continues to infinity along the symmetry axis to satisfy the boundary condition).

\section{General Line-of-sight projection}
\label{sec:los_projection}

When the Jeans equations are used to study the intrinsic kinematics of galaxies (e.g. from Gaia data), or when they are used to compute the starting conditions for the particles of N-body models \citep[e.g.][]{Emsellem2013}, a solution of the equations in \autoref{sec:spherical_jeans_solution} is all that is needed. However for most of the galaxies in the Universe, currently, only projected quantities can be observed. In this situation, one has to project the kinematic along the line-of-sight (LOS) to compute a prediction of the model observables to compare with the observations.

A list of formulas for the projection of an axisymmetric model in cylindrical coordinates was given e.g. in Appendix~A of \citet{Evans1994}. However, I have not found a similar treatment for the spherically-aligned case. The only expression I found is equation~(8) of \citet{Bacon1985} for the second moment of the line-of-sight velocity. However, that expression misses one term and is only correct in the semi-isotropic case. For these reasons, instead of merely listing the final formulas, I give a concise tutorial about the general procedure for the derivation of the line-of-sight projections here. I additionally provide a compact description, in matrix notation, for the corresponding transformation from cylindrical to sky coordinates.

\subsection{From spherical to sky coordinates}
\label{sec:sph_trans}

I adopt the standard convention of measuring the angle $\theta$ from the $z$-axis and the angle $\phi$ from the $x$-axis, in the $x$--$y$ plane (see \autoref{fig:Spherical_Coordinates}). The components of a vector $(v_r,v_\theta,v_\phi)$ in the spherical-polar basis can be transformed into the components of a vector $(v_x,v_y,v_z)$ in the Cartesian basis as follows \citep[e.g. section~3.10 of][]{Arfken2013mathematical}
\begin{equation}
\begin{pmatrix}
v_x \\ v_y \\ v_z 
\end{pmatrix} 
= \mathbf{R}\cdot
\begin{pmatrix}
v_r \\ v_\theta \\ v_\phi 
\end{pmatrix}
\quad\text{with}\quad
\mathbf{R} = 
\begin{pmatrix}
\sin\theta\cos\phi &  \cos\theta\cos\phi  &  -\sin\phi         \\
\sin\theta\sin\phi  & \cos\theta\sin\phi  &  \cos\phi  \\
\cos\theta &  -\sin\theta    &  0  
\end{pmatrix} 
\end{equation}

I assume the Cartesian system $(x,y,z)$ has the $z$-axis aligned with the galaxy symmetry axis and the $x$-axis aligned with the projected major axis, parallel to the plane of the sky. I define an additional inclined $(x',y',z')$ Cartesian system of coordinates with the $x'$-axis coincident with the $x$-axis and the $z'$-axis parallel to the LOS. I define the inclination $i$ as the angle between $z$ and $z'$, which implies $i=90^\circ$ when the galaxy is edge-on, as in the most common convention. A vector in the galaxy Cartesian system $(x,y,z)$ transforms into the observer's system $(x',y',z')$ as follows
\begin{equation}\label{eq:matrix_los}
\begin{pmatrix}
v_{x'} \\ v_{y'} \\ v_{z' }
\end{pmatrix} 
= \mathbf{S}\cdot
\begin{pmatrix}
v_x \\ v_y \\ v_z
\end{pmatrix}
\quad\text{with}\quad
\mathbf{S} =
\begin{pmatrix}
1 &  0           &  0         \\
0 &  {\cos i} &  {-\sin i}  \\
0 &  {\sin i}    &  {\cos i}  
\end{pmatrix}
\end{equation}
Note that both matrices are orthogonal, namely $\mathbf{R}\cdot\mathbf{R}^T=\mathbf{S}\cdot\mathbf{S}^T=\mathbf{I}$, with $\mathbf{I}$ the identity matrix.
The general rules of transformation of tensors \citep[e.g. section~4.1 of][]{Arfken2013mathematical} now imply that the second order tensor in spherical basis, represented by a $3\times3$ matrix, with zero non-diagonal terms due to the assumed spherical alignment\footnote{Of course the expression is generally valid, even when the velocity ellipsoid is not radially oriented, in which case the initial tensor would not be diagonal.}, transforms into a symmetric tensor in the observer's Cartesian basis as
\begin{align}\label{eq:second_moment__projection}
\mathbf{T}=
\begin{pmatrix}
\overline{v_{x'}^2} &  \overline{v_{x'}v_{y'}} &  \overline{v_{x'}v_{z'}}  \\
\overline{v_{y'}v_{x'}} &  \overline{v_{y'}^2} &  \overline{v_{y'}v_{z'}}  \\
\overline{v_{z'}v_{x'}} &  \overline{v_{z'}v_{y'}}    &  \overline{v_{z'}^2}  
\end{pmatrix}
=\mathbf{Q}\cdot
\begin{pmatrix}
\overline{v_r^2} &  0           &  0 \\
0 &  \overline{v_\theta^2} &  0  \\
0 &  0    &  \overline{v_\phi^2}  
\end{pmatrix}
\cdot\mathbf{Q}^T
\end{align}
with the orthogonal matrix $\mathbf{Q}=\mathbf{S}\cdot\mathbf{R}$
\begin{equation}\label{eq:q} 
\mathbf{Q} = 
\begin{pmatrix}
\sin\theta \cos\phi & \cos\theta \cos\phi & -\sin\phi \\
\sin\theta \sin\phi \cos i - \cos\theta \sin i 
& \cos\theta \sin\phi \cos i + \sin\theta \sin i 
& \cos\phi \cos i \\
\sin\theta \sin\phi \sin i + \cos\theta \cos i
& \cos\theta \sin\phi  \sin i - \sin\theta \cos i 
& \cos\phi \sin i 
\end{pmatrix}
\end{equation}

The first moment of the velocities transform from the spherical (or cylindrical) basis to the observer's basis like all vectors. Considering that in a steady-state axisymmetric system $\overline{v_r} = \overline{v_\theta} = 0$, the relation is
\begin{equation}\label{eq:velocity_projection}
\begin{pmatrix}
\overline{v_{x'}} \\ \overline{v_{y'}} \\ \overline{v_{z'}}
\end{pmatrix} 
= \mathbf{Q}\cdot
\begin{pmatrix}
0 \\ 0 \\ \overline{v_\phi}
\end{pmatrix}.
\end{equation}

All components of the first velocity moment and the second velocity moment tensor, including the non-diagonal terms, can be obtained straightforwardly from \autoref{eq:second_moment__projection} and \autoref{eq:velocity_projection} and I will not list all the resulting expressions. I give, however, for illustration, the projected velocities and the diagonal elements of the second moment tensor in the observer's coordinates, where $x'$ is parallel to the galaxy projected major axis, $y'$ is parallel to the projected minor axis and $z'$ is along the LOS. This implies that $\overline{v_{\rm los}}\equiv\overline{v_{z'}}$ and $\overline{v^2_{\rm los}}\equiv\overline{v^2_{z'}}$:
\begin{subequations}\label{eq:los_v}
	\begin{align}
	\overline{v_{x'}} &= \overline{v_\phi}\, Q_{13} = -\overline{v_\phi}\, \sin\phi \\
	\overline{v_{y'}} &= \overline{v_\phi}\, Q_{23} = \overline{v_\phi}\, \cos\phi \cos i\\
	\overline{v_{z'}} &= \overline{v_\phi}\, Q_{33} = \overline{v_\phi}\, \cos\phi \sin i.
	\end{align}
\end{subequations}
The elements of the symmetric tensor $\mathbf{T}$ in \autoref{eq:second_moment__projection} can be written as
\begin{equation}
T_{jk} =\, \overline{v_r^2}\, Q_{j1}Q_{k1} + \overline{v_\theta^2}\, Q_{j2}Q_{k2} +  \overline{v_\phi^2}\, Q_{j3}Q_{k3}.
\end{equation}
When the full second velocity moment tensor is needed, this formula is simpler and more efficient for the numerical computation than the following explicit ones. However, as an example, the expressions for the diagonal elements of the second moment tensor are
\begin{subequations}
	\begin{align}
	\overline{v_{x'}^2} =\, & T_{11} 
	= \left(\overline{v_r^2}\, \sin^2\!\theta + \overline{v_\theta^2}\, \cos^2\!\theta\right) \cos^2\!\phi + \overline{v_\phi^2}\, \sin^2\!\phi\\
	\overline{v_{y'}^2} =\, & T_{22} 
	= \overline{v_r^2} \, (\sin\theta \sin\phi \cos i - \cos\theta \sin i )^2 \nonumber\\
	& + \overline{v_\theta^2} \, (\cos\theta \sin\phi \cos i + \sin\theta \sin i)^2
	+ \overline{v_\phi^2}\, \cos^2\!\phi \cos^2\! i\\
	\overline{v_{z'}^2} =\, & T_{33} 
	= \overline{v_r^2} \, (\sin\theta \sin\phi \sin i + \cos\theta \cos i)^2 \nonumber\\
	& + \overline{v_\theta^2}\, (\cos\theta \sin\phi  \sin i - \sin\theta \cos i)^2 
	+  \overline{v_\phi^2}\, \cos^2\!\phi \sin^2\!i.  
	\end{align}
\end{subequations}

\subsection{From cylindrical to sky coordinates}
\label{sec:cyl_to_sky}

The transformation of vectors and tensors from the cylindrical coordinate system to a coordinates system aligned with the plane of the sky and observer's line of sight is completely analogue to what I described in \autoref{sec:sph_trans}. Only the matrix $\mathbf{R}$ is different.

I adopt the standard convention of measuring the angle $\phi$ from the $x$-axis, in the $x$--$y$ plane (see \autoref{fig:Spherical_Coordinates}). The components of a vector $(v_R,v_\phi,v_z)$ in the cylindrical basis can be transformed into the components of a vector $(v_x,v_y,v_z)$ in the Cartesian basis as follows
\begin{equation}
\begin{pmatrix}
v_x \\ v_y \\ v_z 
\end{pmatrix} 
= \mathbf{R}_{\rm cyl}\cdot
\begin{pmatrix}
v_R \\ v_\phi \\ v_z
\end{pmatrix}
\quad\text{with}\quad
\mathbf{R}_{\rm cyl} = 
\begin{pmatrix}
\cos\phi &  -\sin\phi &  0  \\
\sin\phi & \cos\phi   &  0  \\
0          &  0           &  1  
\end{pmatrix} 
\end{equation}

I assume the same Cartesian systems $(x,y,z)$ and $(x',y',z')$ as in \autoref{sec:sph_trans}.
In the case of cylindrical alignment, the transformation of tensors, with zero non-diagonal terms due to the assumed alignment, into a symmetric tensor in the observer's Cartesian basis is
\begin{align}\label{eq:second_moment__projection_cyl}
\begin{pmatrix}
\overline{v_{x'}^2} &  \overline{v_{x'}v_{y'}} &  \overline{v_{x'}v_{z'}}  \\
\overline{v_{y'}v_{x'}} &  \overline{v_{y'}^2} &  \overline{v_{y'}v_{z'}}  \\
\overline{v_{z'}v_{x'}} &  \overline{v_{z'}v_{y'}}    &  \overline{v_{z'}^2}  
\end{pmatrix}
=\mathbf{Q}_{\rm cyl}\cdot
\begin{pmatrix}
\overline{v_R^2} &  0           &  0 \\
0 &  \overline{v_\phi^2} &  0  \\
0 &  0    &  \overline{v_z^2}  
\end{pmatrix}
\cdot\mathbf{Q}_{\rm cyl}^T
\end{align}
with the orthogonal matrix $\mathbf{Q}_{\rm cyl}=\mathbf{S}\cdot\mathbf{R}_{\rm cyl}$, where $\mathbf{S}$ is still given by \autoref{eq:matrix_los}, resulting into
\begin{equation}\label{eq:q_cyl}
\mathbf{Q}_{\rm cyl} = 
\begin{pmatrix}
\cos\phi & -\sin\phi & 0 \\
\sin\phi \cos i  & \cos\phi \cos i & -\sin i \\
\sin\phi \sin i & \cos\phi \sin i & \cos i \\
\end{pmatrix}
\end{equation}

The projection of the first moment of the velocity is the same as for the spherically-aligned case and is still given by \autoref{eq:los_v}. While for the second velocity moment tensor, as an illustration, the resulting expressions for the diagonal elements are
\begin{subequations}
	\begin{align}
	\overline{v_{x'}^2} =\, 
	& \overline{v_R^2}\, \cos^2\!\phi + \overline{v_{\phi}^2}\, \sin^2\!\phi\\
	\overline{v_{y'}^2} =\, 
	& \left(\overline{v_R^2}\, \sin^2\!\phi + \overline{v_{\phi}^2}\, \cos^2\!\phi\right)\cos^2\!i 
	+ \overline{v_z^2}\, \sin^2\!i \\
	\overline{v_{z'}^2} =\, 
	& \left(\overline{v_R^2}\, \sin^2\!\phi + \overline{v_{\phi}^2}\, \cos^2\!\phi\right)\sin^2\!i + \overline{v_z^2}\, \cos^2\!i
	\end{align}
\end{subequations}
The expression for $\overline{v_{z'}^2}$ has been given many times, starting with \citet{Satoh1980}, while the other components were included in the list by \citet{Evans1994} (in both cases with a different definitions for the coordinate systems than adopted here).

\subsection{Line-of-sight integration}

The observed first or second velocity moments are computed by luminosity-weighting the expressions for the components of the projected first or second velocity moment tensor, given in \autoref{sec:sph_trans} and \autoref{sec:cyl_to_sky}, along the LOS as follows
\begin{subequations}\label{eq:los_integ}
	\begin{align}
	&\Sigma(x',y') =\, \int_{-\infty}^{\infty} \nu \dd z',\label{eq:surf_los}\\
	&\Sigma\,\overline{v_{\alpha}}(x',y') =\, \int_{-\infty}^{\infty} \nu\overline{v_{\alpha}} \dd z'\label{eq:los_integ_v}\\
	&\Sigma\,\overline{v_{\alpha} v_{\beta}}(x',y') =\, \int_{-\infty}^{\infty} \nu\overline{v_{\alpha} v_{\beta}} \dd z'
	\end{align}
\end{subequations}
where $\alpha$ and $\beta$ represent one of the three different components $(x',y',z')$ of the velocity (e.g. $\alpha=z'$ for the mean LOS velocity $\overline{v_{\rm los}}$) or the tensor (e.g. $\alpha=\beta=z'$ for the projected LOS second moment $\overline{v^2_{\rm los}}$). In the case of an MGE surface brightness, the integral of \autoref{eq:surf_los} is analytic and  $\Sigma(x',y')$ is given by \autoref{eq:surf}.

To perform the LOS integration, a given set of sky coordinates $(x',y',z')$ along the LOS is transformed into the galaxy $(x,y,z)$ Cartesian coordinate systems with $\mathbf{S}^{-1}=\mathbf{S}^T$
\begin{equation}
\begin{pmatrix}
x \\ y \\ z
\end{pmatrix} 
= \mathbf{S}^T\cdot
\begin{pmatrix}
x' \\ y' \\ z'
\end{pmatrix}
\end{equation}
the trigonometric functions in \autoref{eq:q} or \autoref{eq:q_cyl} can then be evaluated as (see \autoref{fig:Spherical_Coordinates})
\begin{subequations}
	\begin{align}\label{eq:trig}
	&R^2 = x^2 + y^2  & r^2 = R^2 + z^2 \\    
	&\sin\phi = y/R   & \cos\phi = x/R \\
	&\sin\theta = R/r   & \cos\theta = z/r.
	\end{align}
\end{subequations}

When the object under study is at a small distance and covers a large field of view, one needs to include perspective effects in the LOS integration. The matrix projection of \autoref{eq:matrix_los} should be replaced with a perspective transformation  \citep{vanderMarel2002}.

\subsection{PSF convolution}

For the LOS components, the kinematics is generally affected by the instrumental PSF and the atmospheric seeing. To account for this effect I proceed as in Appendix~A of \citet{Cappellari2008}. The observed mean LOS velocity $[\overline{v_{\rm los}}]_{\rm obs}$ and the second moment $[\overline{v_{\rm los}^2}]_{\rm obs}$ are related to the intrinsic ones by the following relations, where PSF represents a normalized MGE PSF
\begin{subequations}
	\begin{align}\label{eq:conv_surf}
	&\Sigma_{\rm obs} =\, \Sigma \otimes {\rm PSF}\\
	&[\overline{v_{\rm los}}]_{\rm obs} =\,
	\frac{(\Sigma \overline{v_{\rm los}}) \otimes {\rm PSF}}{\Sigma_{\rm obs}}\\
	&[\overline{v_{\rm los}^2}]_{\rm obs} =\,
	\frac{(\Sigma \overline{v_{\rm los}^2}) \otimes {\rm PSF}}{\Sigma_{\rm obs}}.
	\end{align}
\end{subequations}

\section{Multi-Gaussian Expansion formalism}
\label{sec:mge_formalism}

To derive solutions for the Jeans equations I make an explicit choice for the parametrization of the number density of the tracer population and the total density (which can include dark matter and a central black hole). I adopt for {\em both} the MGE parametrization \citep{Emsellem1994,Cappellari2002mge}. Strengths of this approach are its flexibility in reproducing with great detail the surface-brightness of real galaxies, its analytic projection, and the availability of a robust method and a corresponding software implementation\footnote{Available from \url{https://pypi.org/project/mgefit/}} to fit the galaxy photometry in a fully-automated manner \citep{Cappellari2002mge}.

The expressions in this section are written in spherical polar coordinates. They can be converted to cylindrical coordinates using the transformation below, which considers that the angles $\theta$ are measured from the symmetry $z$-axis
\begin{equation}\label{eq:sph_transf}
(R,z) = (r \sin\theta, r \cos\theta)
\end{equation}

\subsection{Tracer surface density or surface brightness}

If the $x'$-axis is aligned with the photometric major axis, the surface brightness $\Sigma$ at the location $(x',y')$ on the plane of the sky can be written as
\begin{equation}\label{eq:surf}
\Sigma(x',y') = \sum_{k=1}^N
{\Sigma_{0k} \exp
	\left[
	-\frac{1}{2\sigma^2_k}
	\left(x'^2 + \frac{y'^2}{q'^2_k} \right)
	\right]},
\end{equation}
where $N$ is the number of the adopted Gaussian components, having peak surface brightness $\Sigma_{0k}$, observed axial ratio $q'_k$ and dispersion $\sigma_k$ along the major axis.

\subsection{Deprojection}
\label{sec:deprojection}

The deprojection of the surface brightness to obtain the intrinsic luminosity density is not unique unless the axisymmetric galaxy is seen edge-on ($i=90^\circ$) \citep{Rybicki1987,Kochanek1996}, and the degeneracy becomes quite dramatic when the galaxy is seen at low inclinations  \citep{Gerhard1996,Romanowsky1997,vandenBosch1997,Magorrian1999}.
The MGE method provides a simple possible choice for the deprojection by {\em assuming} that each projected 2-dim Gaussian is deprojected into an intrinsic 3-dim Gaussian \citep{Monnet1992}. One of the advantages of the MGE method is that one can easily enforce the roundness of the model \citep{Cappellari2002mge}, thus producing realistic densities, which look like real galaxies when projected at any angle. 

However, one should keep in mind that the MGE method, like any other alternative technique, cannot eliminate the mathematical degeneracy of the deprojection. In fact this degeneracy represent one of the major uncertainties in the dynamical modelling \citep{Lablanche2012}. Regardless of the adopted technique, I cannot overemphasise the relevance of the deprojection degeneracy on the dynamical models. This crucial fact is sometimes ignored when one constructs overly-detailed dynamical models of galaxies that are far from edge-on, without considering that, at low inclination, the recovered stellar density can only crudely represent the true one, and any inferred dynamical quantity will be significantly in error. With this caveat in mind, the deprojected MGE axisymmetric luminous density $\nu$ can be written as
\begin{equation}\label{eq:dens}
\nu(r,\theta) = \sum_{k=1}^N
\nu_{0k} \exp
\left[
-\frac{r^2}{2\sigma_k^2}
\left(\sin^2\theta+\frac{\cos^2\theta}{q_k^2} \right)
\right],
\end{equation}
where the individual components have the same dispersion $\sigma_k$ as in the projected case of \autoref{eq:surf}, and the intrinsic axial ratio of each Gaussian becomes, in the most common axisymmetric oblate case ($q_k<1$)
\begin{equation}
q_k^2=\frac{q'^2_k-\cos^2 i}{\sin^2 i},
\end{equation}
where $i$ is the galaxy inclination ($i=90^\circ$ being edge-on). The expression for the rarely-used axisymmetric prolate case ($q_k>1$) is
\begin{equation}
q_k^2=\frac{\sin^2 i}{1/q'^2_k-\cos^2 i}.
\end{equation}

The total luminosity $L_k$ of each Gaussian must remain unchanged during deprojection and is obtained by integrating the Gaussians, using respectively either the projected \autoref{eq:surf} or the intrinsic \autoref{eq:dens}
\begin{equation}
L_k = 2\pi\, \Sigma_{0k}\sigma_k^2 q'_k = \nu_{0k}\left(\sigma_k\!\sqrt{2\pi}\right)^3 q_k.
\end{equation}
This gives the following relation between the projected peak surface number density of the tracer $\Sigma_{0k}$ of each Gaussian (often approximated with the observed surface brightness in $L_\odot$ pc$^{-2}$), and the corresponding peak intrinsic number density $\nu_{0k}$  (often quoted in $L_\odot$ pc$^{-3}$)
\begin{equation}
\nu_{0 k} = \frac{\Sigma_{0 k} q'_k}{q_k \sigma_k\! \sqrt{2\pi}}.
\end{equation}

\subsection{Mass density}

The total mass density $\rho$ can be generally described by a different set of M Gaussian components
\begin{equation}\label{eq:mass}
\rho(r,\theta) = \sum_{j=1}^M
\rho_{0j} \exp
\left[
-\frac{r^2}{2\sigma_j^2}
\left(\sin^2\theta+\frac{\cos^2\theta}{q_j^2} \right)
\right].
\end{equation}
Throughout this paper I use the $j$-index to indicate the parameters of the MGE Gaussians related to the gravitational potential and the $k$-index to refer to the parameters of the Gaussians describing the luminosity density or the tracer population.
In the self-consistent case the Gaussians in \autoref{eq:mass} are the same as those in \autoref{eq:dens} and one has $M=N$, $\sigma_j=\sigma_k$, $q_j=q_k$ and $\rho_{0j}=\Upsilon \nu_{0k}$, where $\Upsilon$ is the mass-to-light ratio, which can account for the stellar population and the possible dark matter contribution. In the non-self-consistent case the density does not follow the luminosity. For example it can be described with the sum of two sets of Gaussians: the first derived by deprojecting the surface brightness with \autoref{eq:dens}, and the second e.g. obtained by fitting a (one-dimensional) MGE model to some adopted analytic parametrization for the dark matter \citep*[e.g. NFW,][]{Navarro1996nfw}, or by fitting an estimate of the stellar mass which allows for $M/L$ variations inferred from stellar population models \citep{Mitzkus2017,Li2017imf}.

\subsection{Gravitational potential}

An expression for the gravitational potential generated by the density of \autoref{eq:mass} was given by \citet{Emsellem1994} as a single integral over a finite interval. I used that form in the solution of the cylindrically-aligned Jeans equations in \citet{Cappellari2008}. Here I proceed differently and use instead the original form of the gravitational potential derived with the general formula for densities stratified on similar ellipsoids  (Sec.~20 of \citealt{Chandrasekhar1969}; Sec.~2.3 of of \citealt{Binney1987})
\begin{equation}\label{eq:chandra}
\Phi(R,z) = \pi G q \int_{0}^{\infty} \frac{\dd u}{\Delta(u)} \int_{Q(u)}^{\infty} \rho(m^2)\,\dd m^2,
\end{equation}
where
\begin{align}
m^2 &= R^2 + z^2/q^2\\
\Delta(u) &= (1 + u) \sqrt{q^2 + u}\\
Q(u) &= \frac{R^2}{1 + u} + \frac{z^2}{q^2 + u}.
\end{align}
This is valid for both oblate ($q<1$) and prolate ($q>1$) density distributions. Substituting \autoref{eq:mass} into \autoref{eq:chandra} and performing the analytic inner integral separately for every $j$-th Gaussian gives
\begin{equation}\label{eq:mge_potential}
\Phi(r,\theta) = -2\pi G \int_0^{\infty}\sum_{j=1}^M  \frac{\rho_{0j} q_j \sigma_j^2 \exp\left[-\frac{r^2}{2\sigma_j^2}\left(\frac{\sin^2\theta}{1+u}+\frac{\cos^2\theta}{q_j^2+u}\right)\right]}{(1+u)\sqrt{q_j^2+u}} \dd u.
\end{equation}
Rather than transforming this integral into a finite interval, I deal with the way of performing this semi-infinite integral as an implementation detail, which I discuss in \autoref{sec:implementation}. This allows for testing of alternative approaches and produces a more robust and efficient implementation of the numerical solution.

The circular velocity is often a useful quantity to extract from the models e.g. to describe the motion of the gas in a galaxy equatorial plane ($z=0$). Using the MGE potential above, this is computed at the galactocentric radius $R$ as
\begin{equation}
v_c^2(R) = -R \frac{{\partial \Phi}}{\partial R} 
= 2 \pi G R^2 \int_{0}^{\infty}\sum_{j=1}^M 
\frac{\rho_{0j} q_j \exp\left[-\frac{R^2}{2 \sigma_j^2 (1 + u)}\right]}
{(1 + u)^2 \sqrt{q_j^2 + u}}\dd u.
\end{equation}
This numerical quadrature can be done with the same DE transformation for the $u$ variable used for the gravitational potential in \autoref{sec:integral_tranformation}.

A supermassive black hole can be modelled by adding the analytic Keplerian potential to \autoref{eq:mge_potential} and deriving a specialized simpler Jeans solution. However, I proceed as in \citet{Cappellari2008} by modelling it as as a small Gaussian having mass $M_j=M_\bullet$, $q_j=1$ and $3\sigma_j\lesssim r_{\rm min}$, where $r_{\rm min}$ is the smallest distance from the black hole that one needs to accurately model (e.g.\ one could choose $r_{\rm min}\approx\sigma_{\rm psf}$).

\section{Jeans solution for an MGE model}
\label{sec:mge_jeans_solution}

In this section, I specialize the general spherically-aligned Jeans solution to the case in which both the tracer population and the total mass density distribution are parametrized with an MGE model.

\subsection{Solution for each luminous Gaussian}

Replacing the tracer density $\nu$ of \autoref{eq:dens} and the gravitational potential $\Phi$ of \autoref{eq:mge_potential} into \autoref{eq:Phi} and \autoref{eq:sig_r}, I obtain the radial dispersion for each luminous Gaussian of the MGE as
\begin{align}\label{eq:sig_r_mge1}
[\nu\overline{v_r^2}]_k =\,& 2\pi G\, \int_r^\infty \dd r'\,
\Biggl[
\nu_{0k} \exp(\mathcal{A}_k + \mathcal{B}_k)\, r'(r'/r)^{2\beta_k}  \nonumber\\
&\times\int_{0}^{\infty} \dd u\, \sum_{j=1}^{M} \frac{\rho_{0j} q_j \exp(\mathcal{C}_{j} + \mathcal{D}_{jk})}
{(1+u)\,(q_j^2 + u)^{3/2}} 
\Biggr] 
\end{align}
with
\begin{align}
\mathcal{A}_k& = -\frac{r'^2}{2 q_k^2 \sigma_k^2}& 
\mathcal{B}_k& = \frac{(1 - q_k^2)\,\mathcal{E}_k}{2 q_k^2 \sigma_k^2}\\
\mathcal{C}_{j}& = -\frac{r'^2}{2 (q_j^2 + u)\,\sigma_j^2}& 
\mathcal{D}_{jk}& = \frac{(1 - q_j^2)\,\mathcal{E}_k}{2\, (1 + u)\,(q_j^2 + u)\, \sigma_j^2}\\
\mathcal{E}_{k}& = (r'/r)^{2\beta_k} (r\sin\theta)^2 
\end{align}

Now replacing \autoref{eq:sig_r_mge1} into \autoref{eq:v2_phi} and considering that the only angular dependency in the expression for $[\nu\overline{v_r^2}]_k$ is inside $\mathcal{E}_{k}$, I obtain an expression for the tangential second velocity moment as
\begin{align}\label{eq:v2_phi_mge1}
[\nu\overline{v_\phi^2}]_k = &\, 2\pi G\, (1 - \beta_k) \int_r^\infty  \dd r'\,
\Biggl\{
\nu_{0k} \exp(\mathcal{A}_k + \mathcal{B}_k)\, r'(r'/r)^{2\beta_k} \nonumber \\ 
&\times\int_{0}^{\infty}\!\! \dd u \sum_{j=1}^{M} \frac{\rho_{0j} q_j \left[1 + 2(\mathcal{B}_k + \mathcal{D}_{jk})\right]\, \exp(\mathcal{C}_{j} + \mathcal{D}_{jk})}
{(1+u)\,(q_j^2 + u)^{3/2}}
\Biggr\} \nonumber\\
&+ \nu_k(r,\theta)\, \sum_{j=1}^{M} \frac{\partial \Phi_j(r,\theta)}{\partial \theta}\tan\theta
\end{align}
where $\nu_k(r,\theta)$ is one term of the sum in \autoref{eq:dens} and
\begin{align}\label{eq:dpot}
\frac{\partial \Phi_j(r,\theta)}{\partial \theta}\tan\theta =\, &
2\pi G\, \int_{0}^{\infty} \Biggl\{\frac{\rho_{0j} q_j (q_j^2 - 1) (r\sin\theta)^2}{(1+u)\,(q_j^2 + u)^{3/2}}\nonumber\\
& \times\exp\left[-\frac{r^2}{2\sigma_j^2}\left(\frac{\sin^2\theta}{1+u}+\frac{\cos^2\theta}{q_j^2+u}\right)\right]\Biggl\} \dd u.
\end{align}
In a more compact form \autoref{eq:sig_r_mge1} and \autoref{eq:v2_phi_mge1} can be rewritten as
\begin{subequations}\label{eq:jeans_sol_compact}
	\begin{align}
	[\nu\overline{v_r^2}]_k =\, & \, 2\, \pi\, G\, \int_r^\infty\!\!\!\! \int_{0}^{\infty}\! \sum_{j=1}^{M} \mathcal{F}_{jk}\, \dd u \dd r'\label{eq:sig_r_mge2}\\
	[\nu\overline{v_\phi^2}]_k =\, & \, 2\, \pi\, G\, (1 - \beta_k)\!\int_r^\infty\!\!\!\! \int_{0}^{\infty}\! 
	\sum_{j=1}^{M}\left[1 + 2(\mathcal{B}_k + \mathcal{D}_{jk})\right]\, \mathcal{F}_{jk}\,
	\dd u \dd r' \nonumber\\
	&+ \nu_k\, \sum_{j=1}^{M} \frac{\partial \Phi_j}{\partial \theta}\tan\theta
	\label{eq:v_psi_mge2}
	\end{align}
\end{subequations}
with
\begin{align}\label{eq:Fjk}
\mathcal{F}_{jk} = 
\frac{\nu_{0k}\, \rho_{0j}\, q_j\, \exp(\mathcal{A}_k + \mathcal{B}_k + \mathcal{C}_{j} + \mathcal{D}_{jk})\, r'(r'/r)^{2\beta_k}}
{(1+u)\,(q_j^2 + u)^{3/2}}.
\end{align}

The outer $r'$ integral in \autoref{eq:jeans_sol_compact} can be written analytically when $2\beta_k$ is integer. The outer integral can also be written in terms of special functions along the symmetry axis $\theta=0$. But these special cases are of little usefulness in practice, so I won't write down the relevant expressions.

In the semi-isotropic limit $\beta_k=0$ the spherically-aligned MGE Jeans solution coincides with the cylindrically-aligned one, given as a single quadrature in \autoref{eq:sigma_z} and \autoref{eq:v2_phi_cyl}. And in the spherical limit, the solution coincides with the spherical one given as single quadrature in \autoref{eq:spherical_jeans}. Moreover, when $\beta_k=\beta$ is constant for the different MGE Gaussians, the inner $u$ integral in \autoref{eq:sig_r_mge1} does not depend on $k$, allowing for a potential speedup of the calculation.

\subsection{Solution for the whole MGE model}

After computing the $[\nu \overline{v_r^2}]_k$ and $[\nu \overline{v_\phi^2}]_k$ solutions, the intrinsic velocity dispersion components and the mean streaming motion of the whole MGE are then computed as
\begin{subequations}
	\begin{align}
	&\nu\sigma^2_r = \nu\overline{v_r^2} = \sum_{k=1}^N [\nu \overline{v_r^2}]_k\\
	&\nu\sigma^2_\theta = \nu\overline{v_\theta^2} = \sum_{k=1}^N (1 - \beta_k)\,[\nu \overline{v_r^2}]_k\\
	&\nu\sigma_\phi^2 = \sum_{k=1}^N (1 - \gamma_k)\,[\nu \overline{v_r^2}]_k\label{eq:sig_phi}\\
	&\nu\overline{v_\phi^2} = \sum_{k=1}^N [\nu \overline{v_\phi^2}]_k\\
	&\nu\overline{v_\phi}^2 = \nu\overline{v_\phi^2} - \nu\sigma_\phi^2\label{eq:vphi}
	\end{align}
\end{subequations}

The Jeans equations do not constrain the splitting of $\overline{v_\phi^2}$ into ordered $\overline{v_\phi}$ and random $\sigma_\phi$ motions. This can be understood physically from the fact that, for a given equilibrium model, one can always revert the sense of rotation of an arbitrary set of orbits, without affecting neither the $\overline{v_\theta^2}/\overline{v_r^2}$ ratio, nor the gravitational potential, nor the distribution of the tracer population. This statement is the anisotropic analogue of the result that, in two-integral, semi-isotropic models, the density distribution determines only the part of the DF that is even in the axial angular momentum \citep{Lynden-Bell1962}. 

For this reason, the splitting of $\overline{v_\phi^2}$ can be performed in an arbitrary way of which \autoref{eq:sig_phi} only represents a possible choice\footnote{The \autoref{eq:vphi} specifies the magnitude of $\overline{v_\phi}$ but not its direction. To model counter-rotating stellar components one can adopt a different velocity sign for the different MGE Gaussians \citep[e.g. fig.~12 of][]{Cappellari2016}.}. Another simple alternative is to use the approach first proposed by \citet{Satoh1980} in the isotropic case and also adopted e.g. by \citet{Binney1990} and \citet{vanderMarel1990}. In that case, it consists of assuming the velocity field $\overline{v_\phi}$ is a scaled version of that of the isotropic model, for which $\sigma_R=\sigma_\phi=\sigma_z$. The analogue assumption, for the cylindrically-aligned anisotropic case, was used in \citet{Cappellari2008} as it appears to describe well real observations \citep[see review by][]{Cappellari2016}. It assumes the velocity field is a scaled version of that of a model with {\em oblate} velocity ellipsoid, for which $\sigma_R=\sigma_\phi\neq\sigma_z$. 

When using the analogue of \citet{Satoh1980} approach, given the spherical symmetry of the alignment adopted here, there are two natural possibilities for the reference model used to define the shape of the $\overline{v_\phi}$: (i) either to assume a model with velocity ellipsoid axially symmetric around the radial $r$-axis, namely $\sigma_r\neq\sigma_\theta=\sigma_\phi$. This choice satisfies the symmetry requirement along the symmetry $z$-axis and naturally converges to a non-rotating spherically-symmetric model in the spherical limit. (ii) Alternatively, one can assume a model with symmetry around the $\theta$ direction, namely $\sigma_r=\sigma_\phi\neq\sigma_\theta$. This model has an oblate velocity ellipsoid in the equatorial plane, but looks unrealistic near the symmetry axis, or in the spherical limit. These two choices imply respectively
\begin{align}\label{eq:vphi2_split_satoh}
[\overline{v_\phi}]_k &= \kappa_k \left[[\overline{v_\phi^2}]_k - (1 - \beta_k)[\overline{v_r^2}]_k\right]^{1/2}\\
[\overline{v_\phi}]_k &= \kappa_k \left([\overline{v_\phi^2}]_k - [\overline{v_r^2}]_k\right)^{1/2}.
\end{align}
Note that these Satoh-like assumptions do {\em not} imply that the velocity ellipsoid is  itself actually axisymmetric! In all cases, this is only true if $\kappa_k=1$. Instead, in general, once $[\overline{v_\phi}]_k$ is obtained, the corresponding $\sigma_\phi$ is given implicitly by \autoref{eq:vphi}. Unlike the assumption of \autoref{eq:sig_phi}, these Satoh-like assumptions generally correspond to a $\gamma_k$ anisotropy that varies spatially even for each single Gaussian component.

\section{Numerical implementation}
\label{sec:implementation}

\begin{figure}
	\includegraphics[width=\columnwidth]{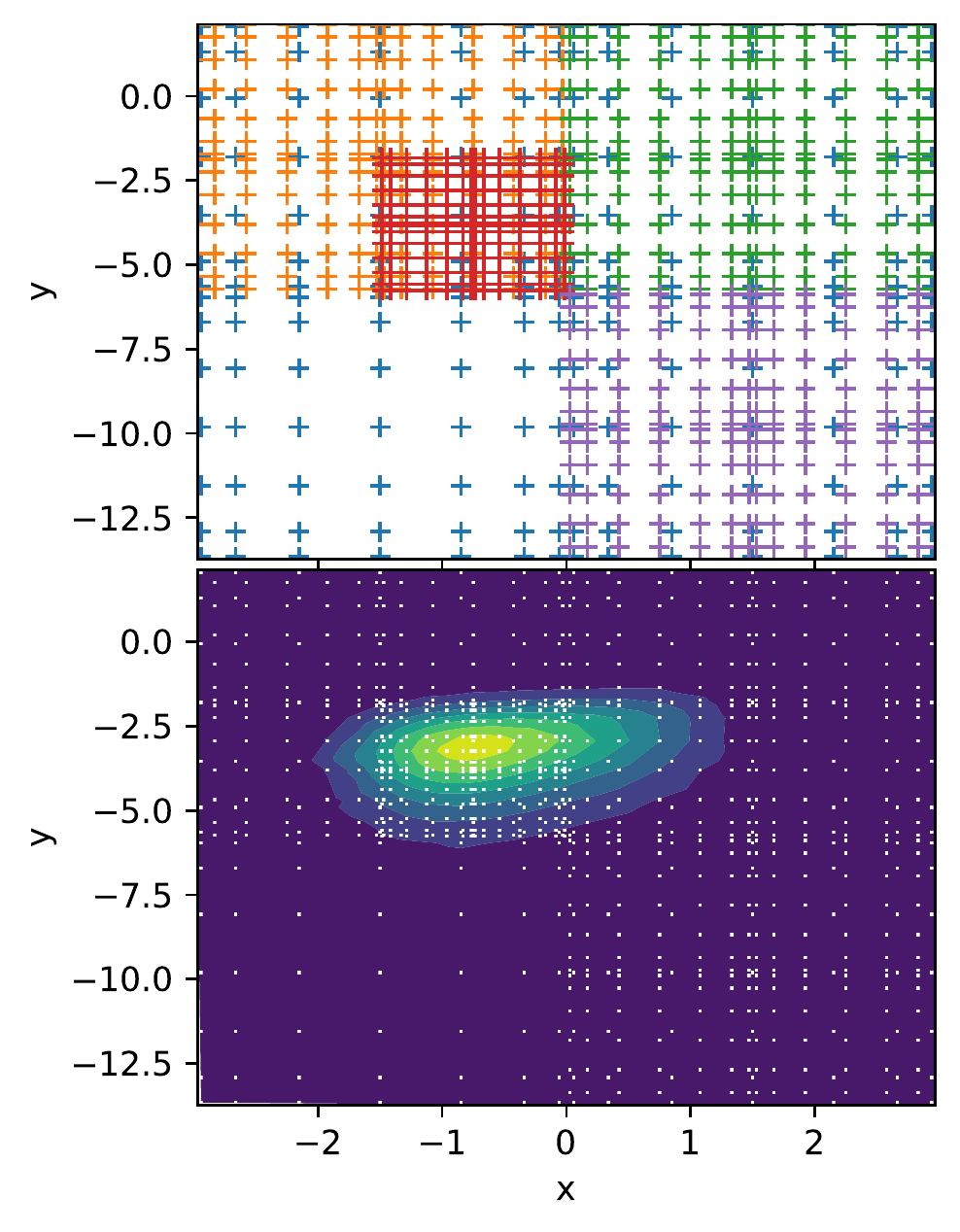}
	\caption{The top panel shows with crosses of different colours the function evaluations at different stages of the refinement process of the adaptive two-dimensional quadrature, where denser crosses imply later stages. The two-dimensional integrand is only evaluated densely where the corresponding sub-integral is not sufficiently accurate. The bottom panel shows the contours of the integrand of \autoref{eq:v_psi_mge2} with over-plotted all locations where it was evaluated. Here the $x$-axis is the $u$ coordinate mapped onto the $x\in[-3,3]$ interval with a DE transformation, and the $y$-axis is $r'$ coordinate mapped to the interval $y\in\ln([10^{-6},r_{\rm max}])$ with a TANH transformation (see \autoref{sec:integral_tranformation} for an explanation).\label{fig:quad2d}}
\end{figure}

The numerical evaluation of the intrinsic first and second velocity moments of \autoref{sec:mge_jeans_solution} requires two nested quadratures, while an additional nested quadrature is needed for the LOS integration of \autoref{eq:los_integ}. The relevant integrals are improper with semi-infinite intervals and can present sharp peaks for certain sets of parameters. For these reasons, a brute-force approach to this triple quadrature, e.g. as an iterated one-dimensional quadrature, would lead to either an unreliable or a very time-consuming and impractical algorithm.

The efficiency of the numerical computation I describe in this section depends on three implementation choices: (i) the use of a specific two-dimensional adaptive quadrature to limit the increase of the function evaluations with the number of dimensions, (ii) the use of efficient transformations fo the improper semi-infinite integrals and (iii) the exploitation of the axisymmetry of the problem in the LOS integration. I discuss each of these in turn in this section.

\subsection{Two-dimensional adaptive quadrature}

After exploring various alternatives, my approach to evaluating the two integrals of \autoref{eq:jeans_sol_compact} is to treat it as a single two-dimensional integral, which I compute with the specific adaptive two-dimensional quadrature method by \citet{Shampine2008quad2d}, which I implemented in my function \textsc{quad2d} in the Python language \citep{vanRossum1995python}. Apart from its high efficiency, the method is designed to be used with vectorized functions, making optimal use of the Numpy package \citep{Numpy2007} characteristics, or for parallel evaluation by multiple CPU cores. The integrator is based on a pair of quadrature rules by \citet{Kronrod1965} which consists of a 3 point Gaussian formula of a degree of precision 5 embedded in a 7 point formula of a degree of precision 11. 

A graphical illustration of how the adaptive quadrature can reduce the number of function evaluations for the Jeans solution is given in \autoref{fig:quad2d}. The figures show that one achieves a large saving in function evaluations by restricting the refinement of the evaluation coordinates to a small region in the domain. This efficiency would not be possible with the more straightforward approach of using two nested one-dimensional quadratures. The figure also shows how the function rapidly drops to zero before reaching the edges of the integration domain, thanks to the integration transformation discussed in the next section.

\subsection{Choice of transformation for improper integrals}
\label{sec:integral_tranformation}

The integrals of \autoref{eq:jeans_sol_compact} are improper as they have semi-infinite intervals and the standard approach to deal with this situation is by using a variable transformation \citep[e.g. Sec.~4.4 of][]{Press2007}. This changes the improper integral, assumed convergent, into a proper one over a finite interval as follows
\begin{align}
\label{eq:integrand_transformation}
I = \int_{0}^{\infty} f(x) \dd x = \int_{a}^{b} f\left[\phi(t)\right]\, \phi'(t) \dd t \\
\text{with} \quad x=\phi(t) \quad \phi(a)=0 \quad \phi(b)=\infty \nonumber.
\end{align}
I experimented with different semi-infinite transformations like $x=-\log t$, $x=t/(1 - t)$ \citep[e.g. Chapter~3 of][]{Davis1984quadrature}, $x=[t/(1 - t)]^2$ \citep{Shampine2008quadva}, the transformation $x=(1-t^2)/t^2$ originally used for the MGE potential by \citet{Emsellem1994}, the semi-infinite TAHN transformation $x=\exp(t)$ \citep{Schwartz1969tahn_quadrature}, the popular double-exponential DE transformations $x=\exp(\pi/2\sinh t)$ and the corresponding version for exponentially-declining integrands $x=\exp[t-\exp(-t)]$ \citep{Takahasi1974de_quadrature}. The different approaches all provided consistent results within the requested accuracy, albeit with significant variations in the smoothness of the transformed integral and correspondingly different execution times. Ultimately I found the best results experimentally, guided by some theoretical insights, namely by measuring the smallest number of function evaluations for different transformations at a fixed prescribed accuracy, and by studying the behaviour of the transformed integrand at different spatial positions using plots like \autoref{fig:quad2d}, for a variety of realistic test cases evaluating \autoref{eq:jeans_sol_compact}. 

The inner Chandrasekar's integrand in $u$ decreases relatively slowly at large radii like $I\propto u^{-5/2}$ as $u\rightarrow\infty$. This explains the fact that I measured the best performance using the full DE transformation $u=\exp(\pi/2\sinh t)$ with $t\in[-3,3]$. Instead, the outer integrand in $r'$ from the Jeans solution decreases exponentially as $I\propto\exp(-r'^2)$ as $r'\rightarrow\infty$, and is not singular at the lower $r'$ bound. A single exponential is sufficient to effectively achieve DE decrease of the integrand at infinity. This explains why I measured best performance with the TANH transformation $r'=r+\exp(t)$ with $t\in\ln([10^{-6},r_{\rm max}])$, where $r_{\rm max}=3\max(\sigma_1,\cdots\sigma_N)$ is the radius beyond which the MGE surface brightness, and the integrand, become negligible.
Importantly, to make the efficiency of my algorithm insensitive to the scaling of the input, I scale the spatial coordinates and the MGE parameters by requiring ${\rm mean}(\sigma_1,\cdots,\sigma_N)=1$, before calling the integrator.

I computed the single integral of \autoref{eq:dpot} with the one-dimensional adaptive algorithm of \citet{Shampine2008quadva}, which I also ported to Python and is the same I used in the cylindrically-oriented Jeans solution \citep{Cappellari2008}. Also for this improper integral over a semi-infinite interval I used the same $x=\exp(\pi/2\sinh t)$ DE transformation as for the Chandrasekhar's integrand in the two-dimensional ones, as they both have the same asymptotic behaviour.

\subsection{Exploiting axisymmetry in the LOS integration}

For the LOS integration of \autoref{eq:los_integ} I used a different approach. Instead of performing a brute-force quadrature in the additional $z'$ dimension, I exploit the axisymmetry of the problem and in particular the fact that the Jeans solution is independent of $\phi$. I evaluate the model's predictions of \autoref{eq:jeans_sol_compact} only in the meridional $(R,z)$ plane, on a grid which is linear in the logarithm of the elliptical radius $m^2=R^2+(z/q)^2$ and in the eccentric anomaly $E$. This is achieved by defining a logarithmically-spaced radial grid $R_j$ and then computing the moments at the cylindrical coordinate positions $(R,z)=(R_j\cos E_k,  q\, R_j \sin E_k)$, for linearly spaced $E_k$ values in the $[0,\pi/2]$ interval, with $q$ a characteristic (e.g. the median) observed axial ratio of the MGE model. During the computation of the integrals of \autoref{eq:los_integ}, the Jeans solution is simply linearly interpolated from the grid. This makes the computation time of the extra LOS quadrature essentially negligible compared to the double integral.

Also for the improper LOS infinite integral in $z'$ it is efficient to use a variable transformation. Also in this case, the integrand decreases exponentially as $I\propto\exp(-z'^2)$ as $z'\rightarrow\infty$. To achieve a DE decrease of the integrand, a single exponential transformation is needed. For this reason I use the TAHN transformation $x=\sinh t$ for the $(-\infty,\infty)$ interval  \citep{Schwartz1969tahn_quadrature}. After some experimentation, here I scale the variable $t$ in such a way that the break $t=\pm1$ between the linear and exponential regimes of the $\sinh t $ function happens for $x=\pm r_{\rm max}/8$. I also limit the LOS integral to the interval $(-r_{\rm max},r_{\rm max})$ outside which the model surface brightness is negligible. 

\subsection{Availability}

A reference implementation for the spherically-aligned JAM$_{\rm sph}$ method is included in the JAM \citep{Cappellari2008} Python software package\footnote{Available from \url{https://pypi.org/project/jampy/}}  \textsc{jampy} starting from version 6.0. JAM$_{\rm sph}$ complements the cylindrically-aligned JAM$_{\rm cyl}$ and spherical solutions, which were already included in earlier versions of \textsc{jampy}. For all assumed orientations of the velocity ellipsoid, \textsc{jampy} can compute either the intrinsic first or second velocity moments (e.g. to model Milky Way surveys like Gaia or to generate N-body realizations of galaxies) or any component of the line-of-sight velocity first moments or of the second moments tensor (e.g. to model external galaxies).

\section{Jeans solutions for Satoh's model}
\label{sec:satoh}

In this section I provide two relatively simple test cases for both the spherically-aligned and cylindrically-aligned anisotropic Jeans solutions, using the potential-density pair by \citet{Satoh1980}. In both cases the derived anisotropic Jeans solutions require one quadrature less than my general MGE solution, allowing for a reliability test of the latter. Moreover, the radically different formalism compared to the MGE one provides thorough testing of the relatively-cumbersome equations and implementation as well.

\subsection{Spherically-aligned solution}
\label{sec:satoh_sph}

To test the algorithm it is crucial to compare its result against alternative formulas that provide the solution with fewer numerical quadratures. For this one can use potential-density pairs, namely expressions for which both the density and the corresponding self-consistent gravitational potential can be computed analytically. A convenient and sufficiently realistic expression is provided by the \citet{Satoh1980} potential-density pair, which is given in polar coordinates, with $\theta$ measured from the symmetry axis, by 
\begin{align}
&\Phi(r,\theta)=-\frac{G M}{S}\label{eq:satoh_pot}\\
&\nu(r,\theta) = \frac{b^2 M \left[a S^2+3 \left(S^2-r^2\right) 
	\sqrt{b^2 + (r\cos\theta)^2}\right]}
{4 \pi  S^5 \left[b^2 + (r\cos\theta)^2\right]^{3/2}}\label{eq:satoh_dens}\\
&S^2 = a^2 + 2a \sqrt{b^2 + (r\cos\theta)^2}\, + r^2,
\end{align}
where $M$ is the total mass of the model and $(a,b)$ are scale parameters.
Plugging these density and potential into \autoref{eq:sig_r} gives the radial dispersion for the Jeans equations with spherically-aligned velocity ellipsoid as a single integral
\begin{align}\label{eq:sig_r_satoh}
&\nu\overline{v^2_r} = 
\frac{a b^2 G M^2}{4\pi}
\int_r^\infty
\frac{\left(a + Q\right) \left[(a + 2Q)(a + 3Q) + r'^2\right] r' (r'/r)^{2 \beta}}
{\left[Q\left(a^2+2 a Q + r'^2\right)\right]^4}\dd r'\\
&Q^2 = b^2  + r'^2 - (r'/r)^{2 \beta } (r\sin\theta)^2.
\end{align}
The second moment $\nu\overline{v^2_\phi}$ of the tangential velocity is then obtained using \autoref{eq:v2_phi} with $\nu$ from \autoref{eq:satoh_dens}, $\nu\overline{v^2_r}$ from \autoref{eq:sig_r_satoh} and
\begin{align}
\frac{\partial (\nu\overline{v^2_r})}{\partial \theta}\tan\theta =&
\frac{a b^2 G M^2}{4\pi}
\int_r^{\infty}\dd r' \Bigg\{\frac{(r\sin\theta)^2 r'(r'/r)^{4 \beta}}
{\left[Q\left(a^2 + 2 a Q + r'^2\right)\right]^5}\nonumber\\
&\times\bigg[2a Q\left(53 a^2+69 a Q+30 Q^2\right) + 6Q (6 a + Q) r'^2\nonumber\\
& + \left(a^2+r'^2\right)\left(34 a^2+3 r'^2\right)
+ 4a\left(a^2 + r'^2\right)^2\!\!/Q\bigg]\Bigg\}\\
\frac{\partial \Phi}{\partial \theta}\tan\theta =&
-\frac{a G M (r\sin\theta)^2}{S^3 \sqrt{b^2 + (r\cos\theta)^2}}.
\end{align}
The numerical quadratures for the semi-infinite improper integrals in this section can be performed with the same TANH transformation for the $r'$ variable discussed in \autoref{sec:integral_tranformation}.

\subsection{Cylindrically-aligned solution}
\label{sec:satoh_cyl}

The density distribution of the Satoh model can be written in cylindrical coordinates as
\begin{align}
&\nu(R,z) = \frac{a b^2 M \left[3 Q (a+2 Q)+S^2\right]}{4 \pi  Q^3 S^5}\\
&S^2 = a^2 + 2aQ + R^2 + z^2  \\
&Q^2 = b^2 + z^2,
\end{align}
with the corresponding self-consistent gravitational potential still given by the same expression of \autoref{eq:satoh_pot}.

In the isotropic limit the Jeans solutions for both $\overline{v^2_z}$ and $\overline{v^2_\phi}$ can be written analytically and the resulting expressions where given by \citet{Satoh1980}. The same analytic solution applies to the $\overline{v^2_z}$ component in the cylindrically-aligned case when $\beta_z\neq0$. The general Jeans solution in this case is given by \autoref{eq:jeans_sol_z}, which for the Satoh model, replacing the corresponding density and potential, becomes simply
\begin{equation}
\overline{v^2_z}(R,z) = \frac{G M Q (a+2 Q)}{2 S \left[3 Q (a+2 Q) + S^2\right]}
\end{equation}

The general anisotropic $\beta_z\neq0$ Jeans solution for the tangential velocity second moment $\overline{v_\phi^2}$ is given by \autoref{eq:jeans_sol_R}, which for the self-consistent Satoh model I found can be written in the very simple form
\begin{equation}
\overline{v_\phi^2}(R,z) =  \frac{\overline{v^2_z}(R,z)}{1-\beta_z}\left(
1 - \frac{6R^2}{S^2}
\right)
+ \frac{G M R^2}{S^3}.
\end{equation}

\section{Results}

\subsection{Numerical accuracy}

\begin{table}
	\caption{Parameters for the MGE fit to the intrinsic density of the Satoh model of \autoref{fig:mge_satoh} with total mass $M=1$ and scale $a=b=1$}
	\label{tab:mge}
	\centering
	\begin{tabular}{ccc}
		\hline 
		$\lg\nu_{0k}$ & $\lg\sigma_k$ & $q_k$\\
		$(a^{-3})$  & $(a)$ & \\
		\hline     
		-1.834 & -0.238 &  0.581 \\
		-1.686 & -0.093 &  0.695 \\
		-1.934 &  0.053 &  0.374 \\
		-2.208 &  0.076 &  0.739 \\
		-3.019 &  0.228 &  0.808 \\
		-2.339 &  0.236 &  0.397 \\
		-2.977 &  0.378 &  0.162 \\
		-3.850 &  0.406 &  0.792 \\
		-3.171 &  0.417 &  0.424 \\
		-4.960 &  0.485 &  0.970 \\
		-3.305 &  0.558 &  0.174 \\
		-4.964 &  0.643 &  0.653 \\
		-4.305 &  0.644 &  0.386 \\
		-5.610 &  0.694 &  0.863 \\
		-4.124 &  0.754 &  0.170 \\
		-4.057 &  0.781 &  0.074 \\
		-6.695 &  0.809 &  1.000 \\
		-5.611 &  0.972 &  0.271 \\
		-6.350 &  0.998 &  0.493 \\
		-5.160 &  1.068 &  0.118 \\
		-4.596 &  1.072 &  0.058 \\
		-7.518 &  1.085 &  1.000 \\
		\hline
	\end{tabular} 
\end{table}

\begin{figure}
	\includegraphics[width=\columnwidth]{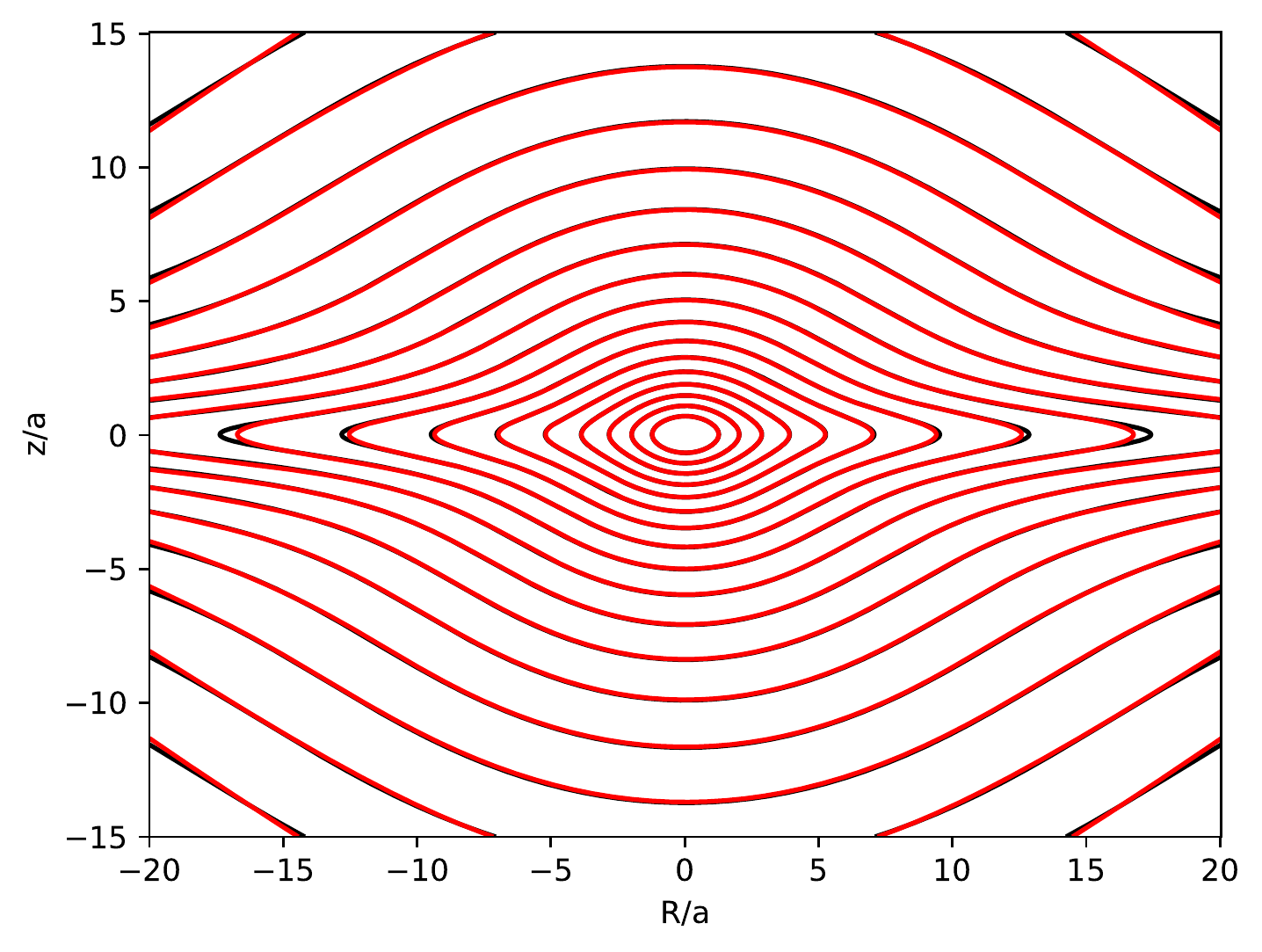}
	\caption{MGE fit to the intrinsic density of a \citet{Satoh1980} model with scale parameters $a=b=1$. The black contours represent the analytic model and the red ones the MGE fit. Contours are spaced by 1 mag.\label{fig:mge_satoh}}
\end{figure}

\begin{figure}
	\includegraphics[width=\columnwidth]{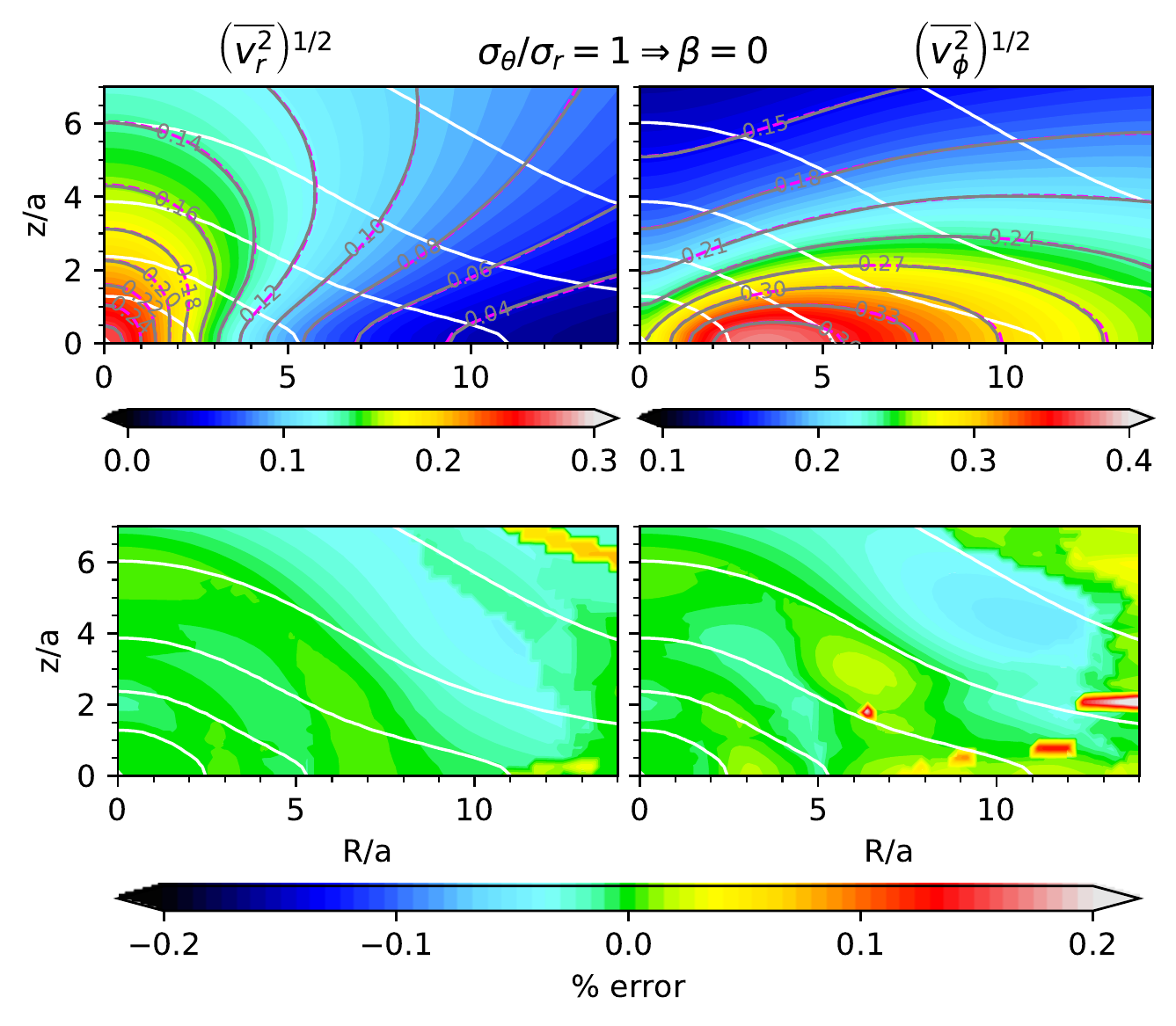}
	\caption{Comparison between the cylindrically-oriented JAM$_{\rm cyl}$ and the new spherically-oriented JAM$_{\rm sph}$ Jeans solutions for the Satoh's model of \autoref{fig:mge_satoh}, in the isotropic limit, where both solutions must be identical. {\em Top Panels:} The colours and the grey contours with labels are the JAM$_{\rm sph}$ solutions while the magenta dashed contours are the solutions of \autoref{sec:satoh_sph}. The unit of velocity is $\sqrt{G M/a}$. The white contours are the model isodensity, spaced by factors of 10 starting from the maximum value. {\em Bottom Panels:} Fractional residuals between JAM$_{\rm sph}$ and JAM$_{\rm cyl}$. In this example, I set an error of 1\% in the adaptive two-dimensional quadrature of JAM$_{\rm sph}$ and a significantly smaller one for the one-dimensional quadrature of JAM$_{\rm cyl}$. The resulting error in JAM$_{\rm sph}$ is always well within the requested accuracy, with small discontinuities dependent on the levels of adaptive refinements employed by the quadrature at a given position.	\label{fig:jam_errors_beta_00}}
\end{figure}

Careful testing is needed to validate the implementation of the equations of \autoref{sec:mge_jeans_solution}. I start by comparing the results for $\overline{v^2_r}$ and $\overline{v^2_\phi}$ of the spherically-aligned Jeans solution against the cylindrically-aligned solution\footnote{I used v6.0 of the \textsc{jampy} package from \url{https://pypi.org/project/jampy/}} of \citet{Cappellari2008} as reproduced in \autoref{eq:v2_phi_cyl} and \autoref{eq:sigma_z}. In the semi-isotropic limit, the velocity ellipsoid is a circle in the meridional plane, which implies that the velocity dispersion is the same along any axis and in particular $\overline{v^2_r}=\overline{v^2_z}$ and the spherically-aligned and cylindrically-aligned solutions must be identical.

For the tests I use as input an MGE fit to the parametrization of the density by \citet{Satoh1980} in \autoref{eq:satoh_dens}, with total mass $M=1$ and scale parameters $a=b=1$. The two-dimensional MGE fit (\autoref{fig:mge_satoh}) was obtained in a fully-automated manner with the method and \textsc{mgefit} Python package\footnote{I used v5.0 of the \textsc{mgefit} package from \url{https://pypi.org/project/mgefit/}} of \citet{Cappellari2002mge}. It consists of 24 Gaussians (\autoref{tab:mge}) and contains 96\% of the total mass of the analytic model. Given that both Jeans solutions use the very same MGE model, but the cylindrically-aligned solution relies on a single quadrature, this test allows me to verify in detail the numerical accuracy of the two-dimensional quadrature. In the computation, I set an accuracy of 1\% on the two-dimensional quadrature  ($\texttt{epsrel}=0.01$ in the procedure \textsc{quad2d}).  The resulting comparison is displayed in \autoref{fig:jam_errors_beta_00}. The maps of residuals show that the accuracy is always well within the requested tolerance, with errors never exceeding 0.2\%. For comparison, the difference between JAM$_{\rm sph}$ and the analytic solution of \autoref{sec:satoh_cyl}, in the semi-isotropic limit, is on the order of a couple of percents, due to the slight differences between the MGE fitted density and the analytic one.

A test of the numerical accuracy for the anisotropic case can be performed in the spherical limit, where the axisymmetric cylindrically-aligned solution converges to the spherical solution of \autoref{sec:mge_spherical}.

\subsection{Intrinsic moments at different anisotropy}
\label{sec:satoh_intr_test}

\begin{figure}
	\includegraphics[width=\columnwidth]{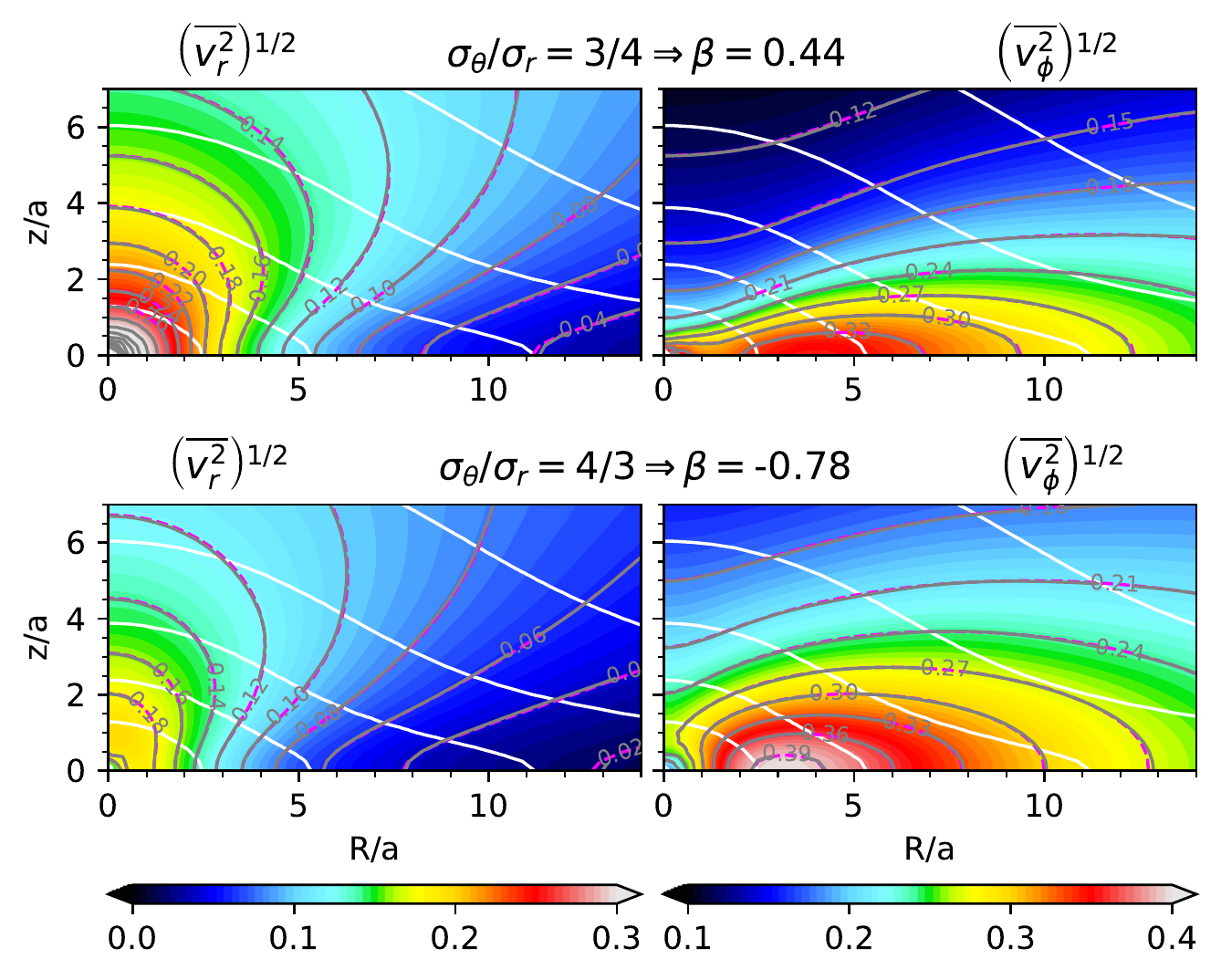}
	\caption{Intrinsic moments of JAM$_{\rm sph}$ for two different anisotropies. The Satoh's model, the meaning of the contour lines and the colour levels are like in the top panels of \autoref{fig:jam_errors_beta_00}. The anisotropy is different and is written in the plot titles.\label{fig:jam_satoh}}
\end{figure}

To test the algorithm in the general anisotropic case, I compare the MGE spherically-aligned Jeans solution presented in \autoref{sec:mge_jeans_solution} against the corresponding solution for the Satoh model presented in \autoref{sec:satoh}. For the tests I used a relatively large anisotropy with axial ratios of the velocity ellipsoid of $\sigma_\theta/\sigma_r=(3/4,1,4/3)$ respectively, corresponding to $\beta=(0.44,0,-0.78)$. The results are shown in \autoref{fig:jam_satoh}. The tests show that the Jeans solution based on the MGE and the one based on the Satoh model agree extremely well. The small differences are because the MGE model does not perfectly reproduce the Satoh density distribution. This is clear from the fact that some differences are also present in the isotropic case, where I know the solution is accurate to the 0.2\% level. The MGE fit could be improved with more Gaussians, but I decided to keep a comparable number of Gaussians as one could use on state-of-the-art photometric observations of real galaxies.

\autoref{fig:jam_satoh} qualitatively illustrates the general trends in the Jeans solution that one should expect to find for real galaxies. Radial anisotropy ($\sigma_r>\sigma_\theta\Rightarrow\beta>0$) produces an increase in both $\overline{v^2_r}$ and $\overline{v^2_\phi}$ towards the centre and a decrease of the tangential component $\overline{v^2_\phi}$ at larger radii. The opposite happens with tangential anisotropy ($\sigma_r<\sigma_\theta\Rightarrow\beta<0$): a central depression develops in both $\overline{v^2_r}$ and $\overline{v^2_\phi}$, while the peak of $\overline{v^2_\phi}$ at larger radii increases. Overall, the mean $\overline{v^2_r}$ decreases while $\overline{v^2_\theta}$ correspondingly increases.

\subsection{Projected moments at different anisotropy}

\begin{figure}
	\includegraphics[width=\columnwidth]{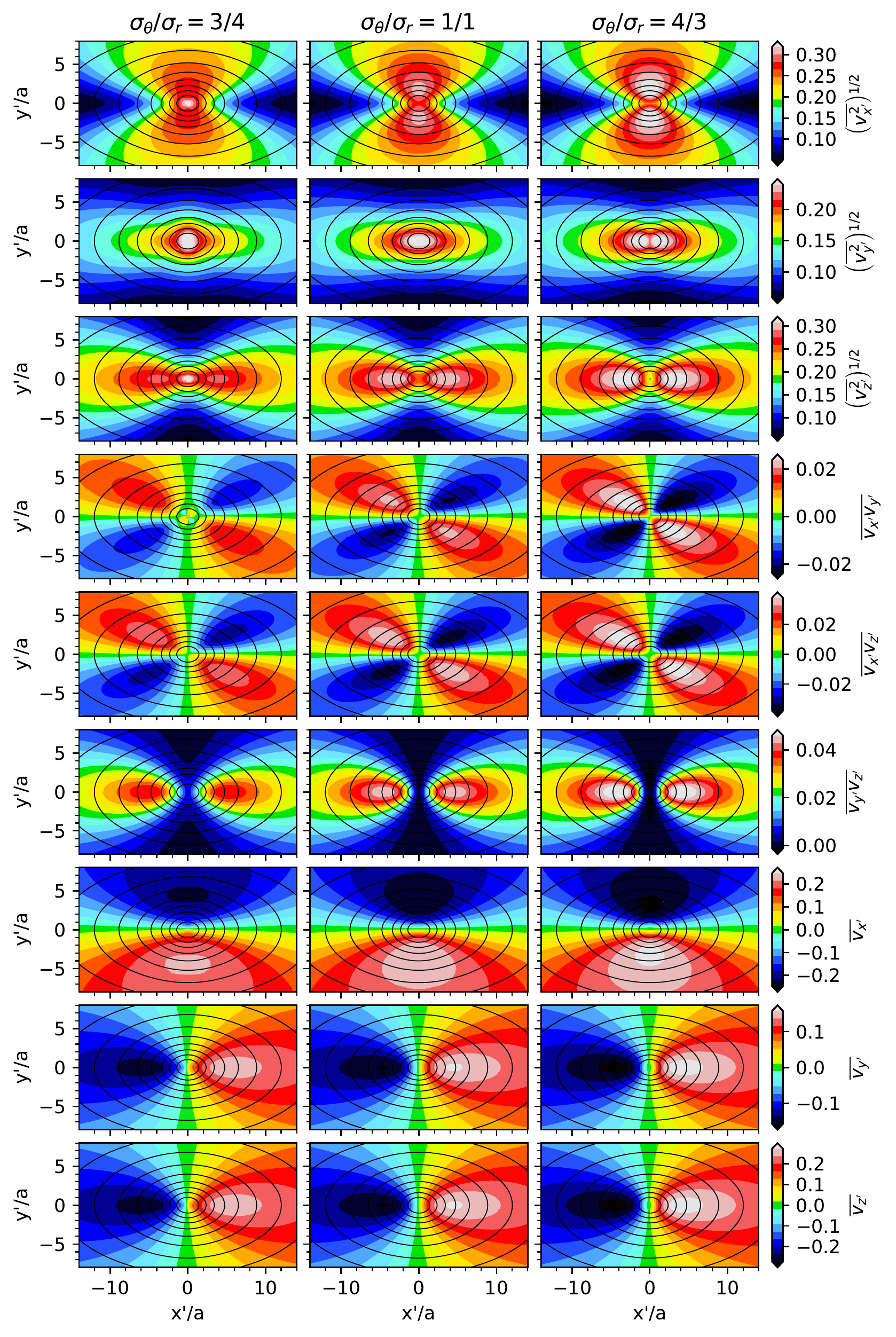}
	\caption{Projected moments for the Satoh's model of \autoref{fig:mge_satoh} seen at an inclination of $i=60^\circ$, for three different anisotropies as written in the titles. The different rows show the six components of the symmetric second velocity moment tensor and the three components of the projected mean velocity, as written in the colour bars. The unit of velocity is $\sqrt{G M/a}$. The black surface brightness contours are spaced by 1 mag. \label{fig:jam_satoh_proj}}
\end{figure}

In \autoref{fig:jam_satoh_proj} I illustrate the qualitative variation of the projected moments as a function of anisotropy, for the same Satoh model as in \autoref{sec:satoh_intr_test}, seen at an inclination of $i=60^\circ$, and the same set of anisotropies as for the intrinsic moments in \autoref{fig:jam_satoh}. The adopted inclination is the average value for random orientations on a sphere. I show all first and second velocity moments, namely the three projected components of the first velocity moment, and all six components of the symmetric second velocity moment tensor. The most easily observable projected moment is the line-of-sight component, namely the mean line-of-sight velocity $\overline{v_{\rm los}}\equiv\overline{v_{z'}}$ and the second line-of-sight velocity moment $\overline{v^2_{\rm los}}\equiv\overline{v^2_{z'}}$. When the kinematics is extracted from observed spectra using a Gaussian approximation for the line-of-sight velocity distribution \citep[e.g.][]{Cappellari2017}, the first moment is empirically approximated by the location of the Gaussian peak $V$ and the second moment by the $V_{\rm rms}^2\equiv V^2 + \sigma^2$, where $\sigma$ is the Gaussian dispersion. 

As discussed in sec.~3.1.5 of \citet{Cappellari2008}, when one is interested in studying mass distributions, one should only fit the second moments and ignore the first ones. This is because the first moments do not contain extra information on the gravitational potential that is not already contained in the second ones. Moreover, the second moments only require an assumption on the $\sigma_\theta/\sigma_r$ ratio and not the $\sigma_\phi/\sigma_r$ one. The first moment also have the issue that one has to split the $\overline{v^2_\phi}$ into order and random motion using \autoref{eq:v2phi_split} and this can lead to unphysical results when $\overline{v^2_\phi}<\sigma_\phi^2$, for the assumed $\gamma$ anisotropy or Satoh-like $\kappa$ parameter. The same considerations summarized for JAM$_{\rm cyl}$ apply unchanged to this JAM$_{\rm sph}$ solution. In practice, to compute the first moments in \autoref{fig:jam_satoh_proj} I assumed, just for reference, a radially symmetric shape for the velocity ellipsoid, namely $\sigma_\theta=\sigma_\phi\Rightarrow\beta=\gamma$. 

From \autoref{fig:jam_satoh_proj} one can generally see the same features already described for the intrinsic moments in \autoref{fig:jam_satoh}. Again, radial anisotropy produces a central peak in the diagonal second moments $(\overline{v^2_{x'}}, \overline{v^2_{y'}}, \overline{v^2_{z'}})$ and reduces the amplitude of the peak in both the first and second moments at larger radii. A central depression in the second moments appears with tangential anisotropy. In the models shown here, I did not include a supermassive black hole, and I did not model seeing effects, to limit the number of arbitrary parameters to explore. It is well known that the presence of a supermassive black hole, which is expected to be present in all stellar spheroids, qualitatively changes the behaviour of the second velocity moments in the centre, generally producing nuclear peaks for a range of surface brightness profiles \citep{Tremaine1994} and anisotropies.

As a test for the projection of all the first and second velocity moments I used the formulas for the cylindrically-aligned Jeans solution (JAM$_{\rm cyl}$) summarized in \autoref{sec:jam_cyl_proj}. For both approaches, I adopted the isotropic model for which the two solutions must coincide. The JAM$_{\rm cyl}$ provides all the projected second moments with a single quadrature \citep{Cappellari2008,Cappellari2012jam}, and the first moments with a two-dimensional quadrature, as opposed to the three quadratures required for JAM$_{\rm sph}$. I found a close agreement, within the uncertainties of the numerical implementation, between the projected model predictions provided by the two radically-different formalisms and implementations.

\subsection{Spherically versus cylindrically aligned solutions}
\label{sec:sph_vs_cyl}

\begin{figure*}
	\includegraphics[width=0.49\textwidth]{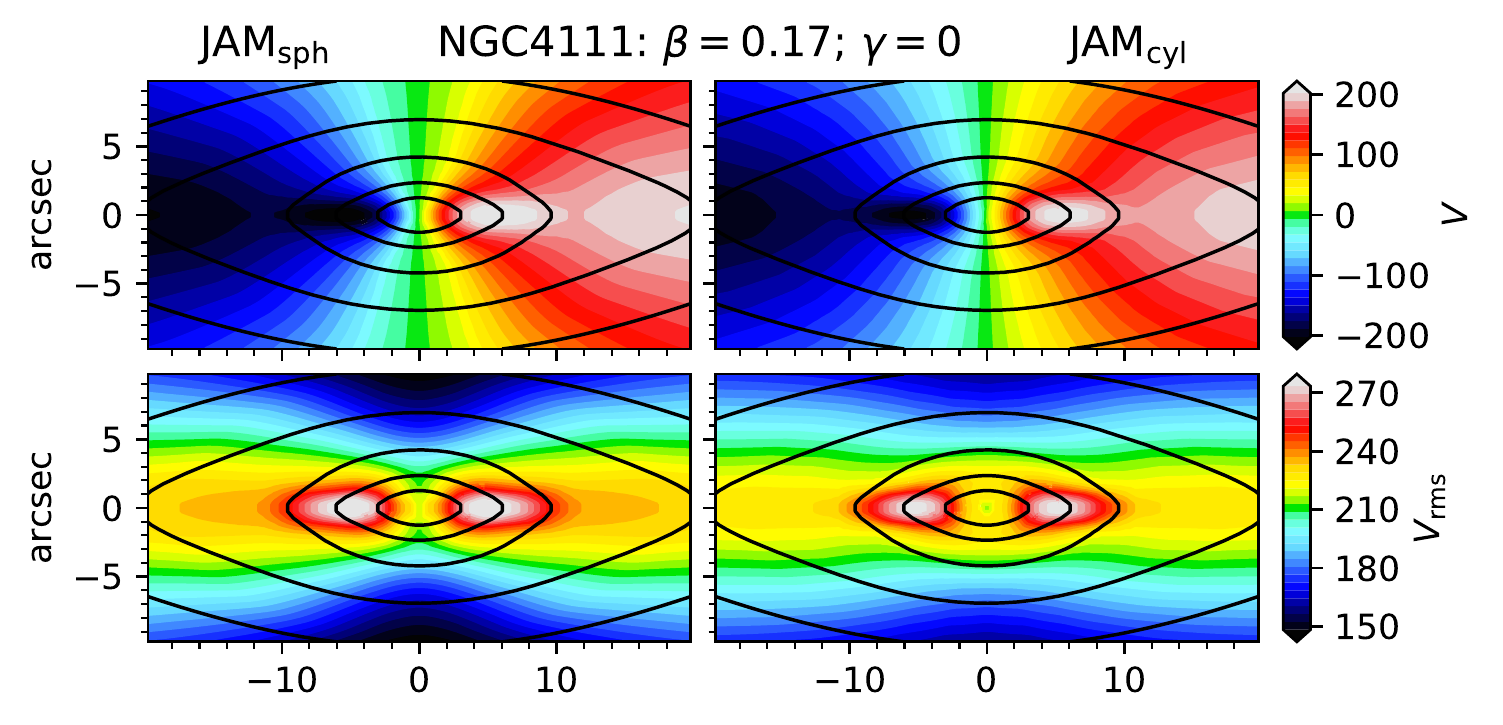}
	\includegraphics[width=0.49\textwidth]{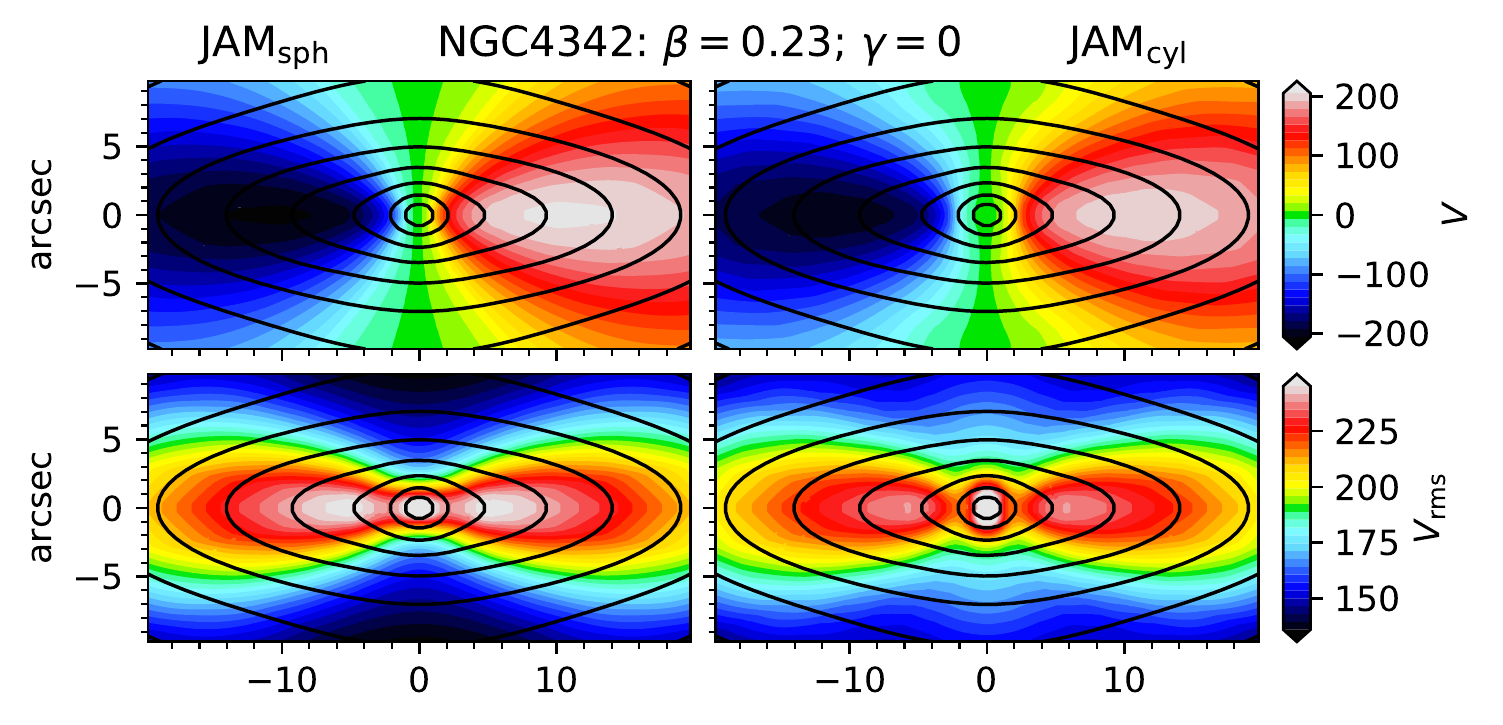}
	\includegraphics[width=0.49\textwidth]{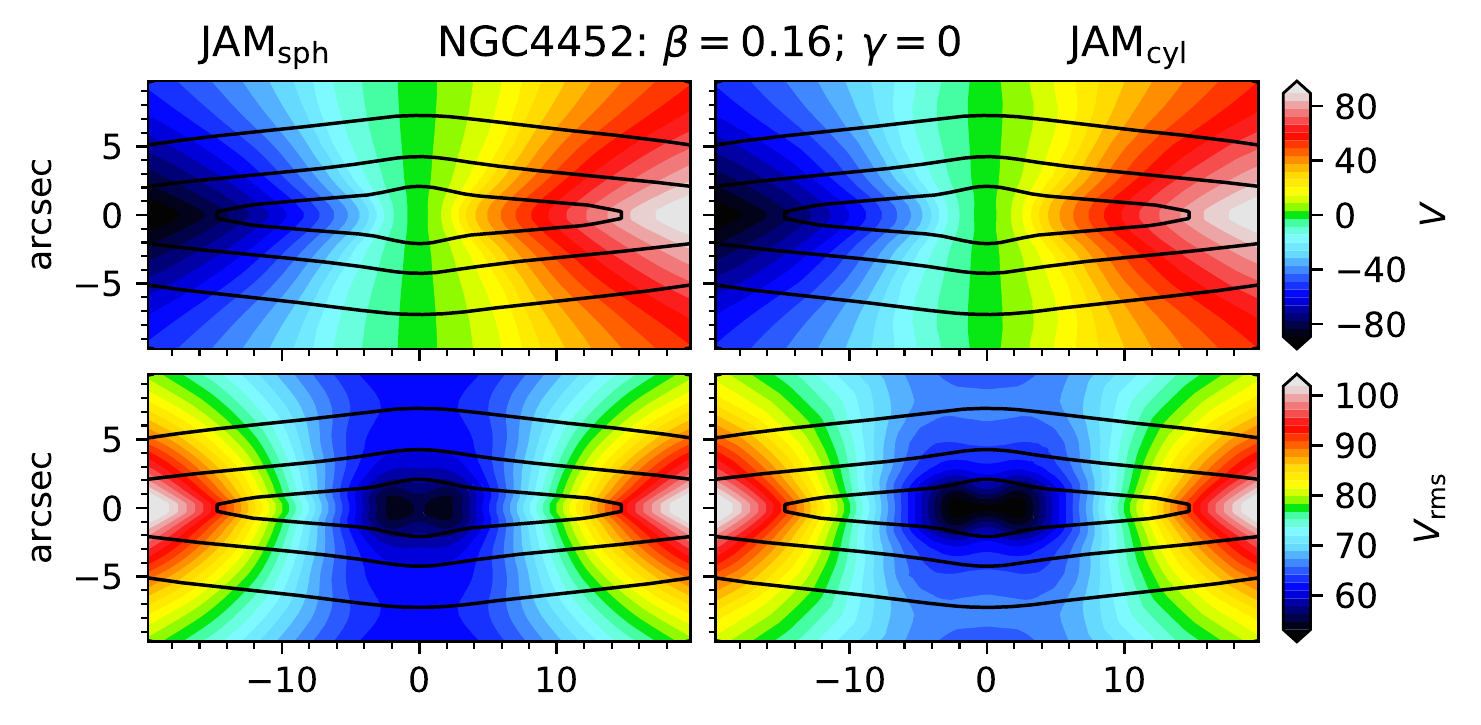}
	\includegraphics[width=0.49\textwidth]{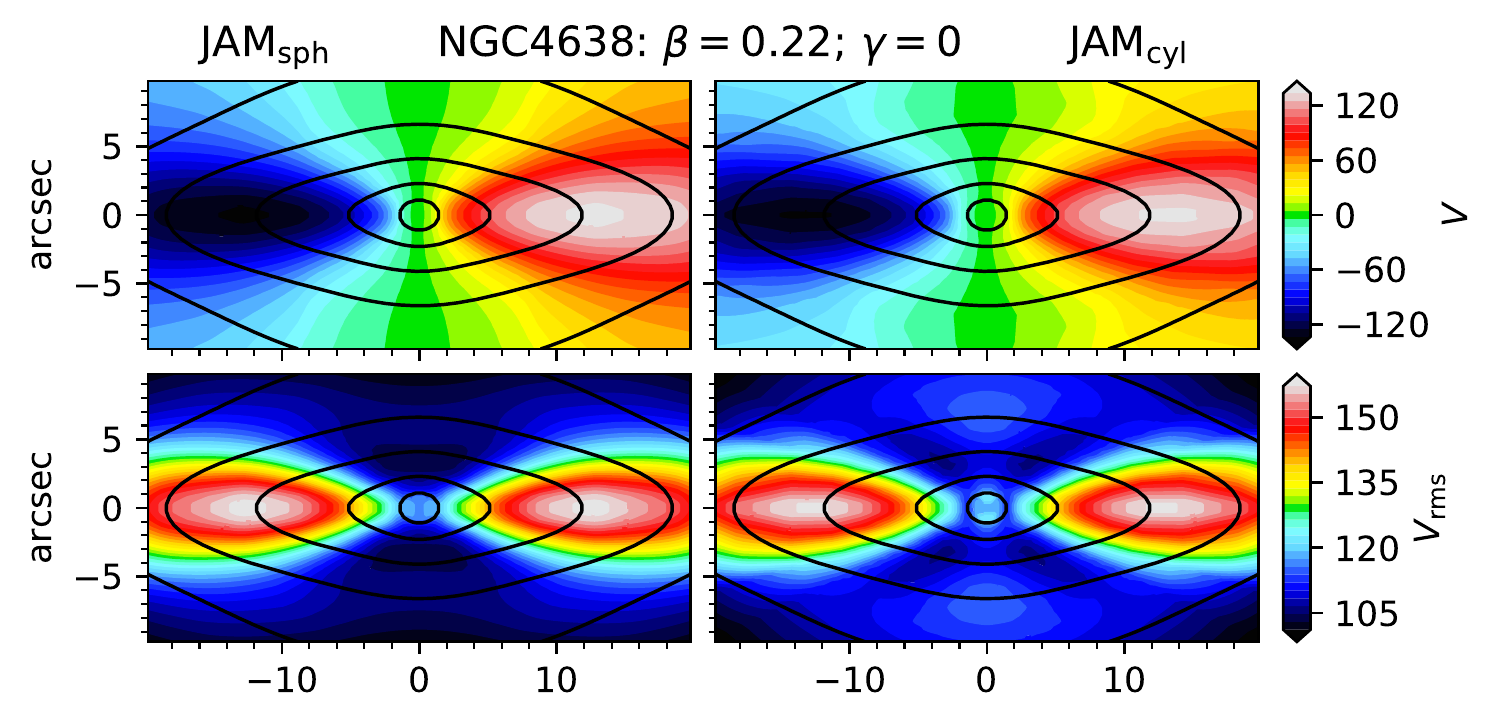}
	\includegraphics[width=0.49\textwidth]{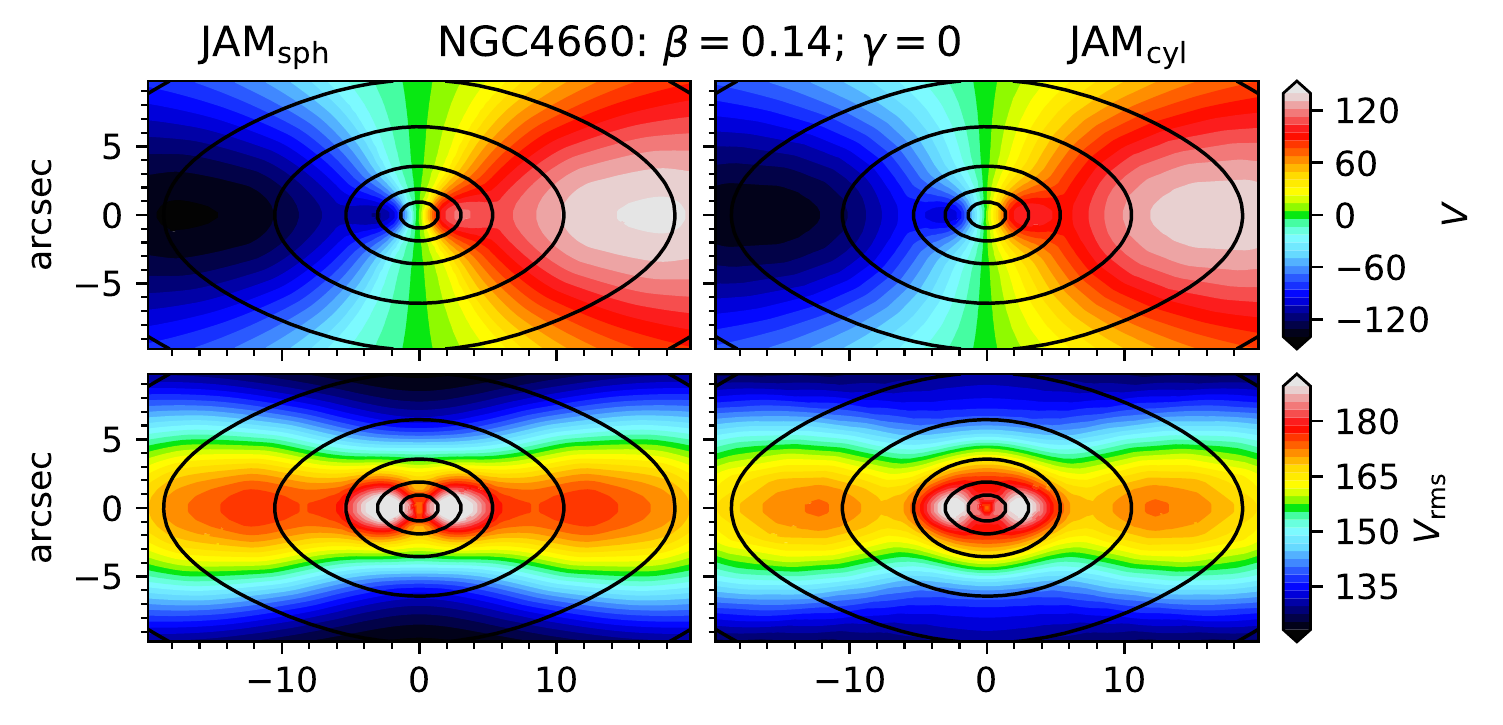}
	\includegraphics[width=0.49\textwidth]{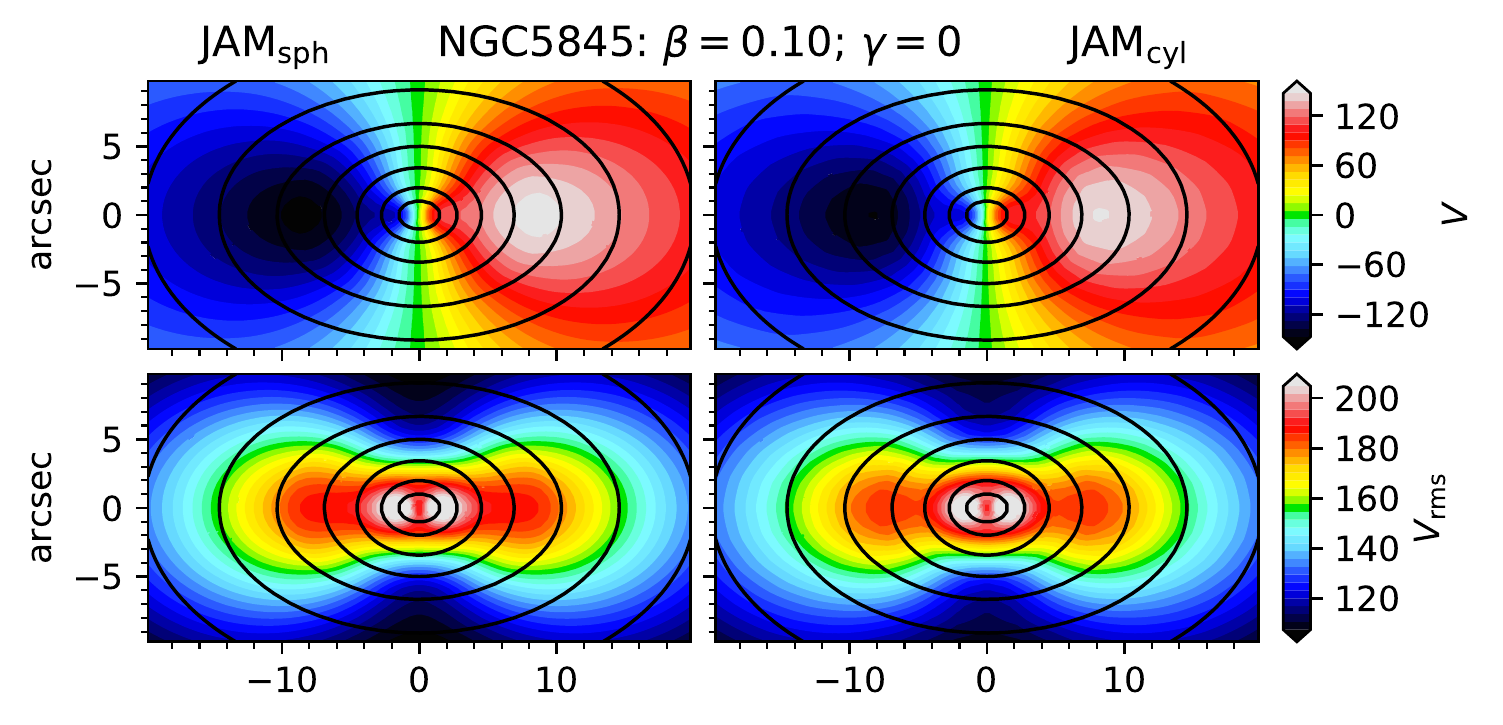}
	\caption{Comparison between the JAM$_{\rm sph}$ (left) and JAM$_{\rm cyl}$ (right) Jeans solutions using the MGEs describing the surface brightness of a set of real galaxies and the corresponding best fitting parameters fitted with JAM$_{\rm cyl}$ to their integral-field kinematics. For each galaxy, the two rows show the mean LOS stellar velocity $V$ and the LOS second velocity moment $V_{\rm rms}$. The $V$ is computed assuming for both models the same shape of the velocity ellipsoid in the equatorial plane (see text for details). The black surface brightness contours are spaced by 1 mag. The kinematics of these galaxies and JAM$_{\rm cyl}$ fits were shown in fig.~10 of \citet{Cappellari2016}. \label{fig:jam_atlas3d}}
\end{figure*}

\begin{figure}
	\centering
	\includegraphics[width=\columnwidth]{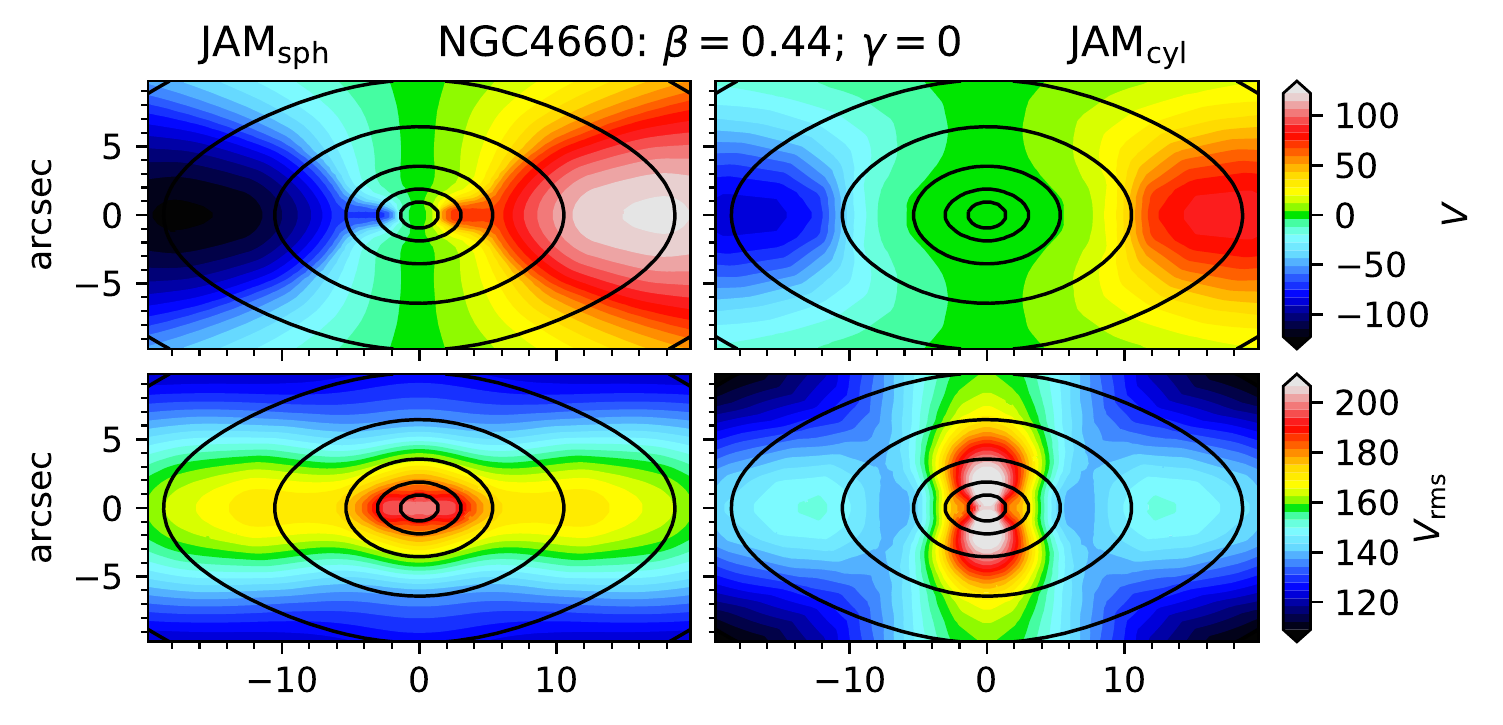}
	\caption{This figure is the same as \autoref{fig:jam_atlas3d}, for the galaxy NGC~4660. Except for the fact that here I adopted an anisotropy $\sigma_\phi/\sigma_r=\sigma_z/\sigma_R=3/4$. Note the strong vertical elongation in the $V_{\rm rms}$ of the JAM$_{\rm cyl}$ solution.\label{fig:jam_ngc4660_beta03}}
\end{figure}

\begin{figure}
	\includegraphics[width=\columnwidth]{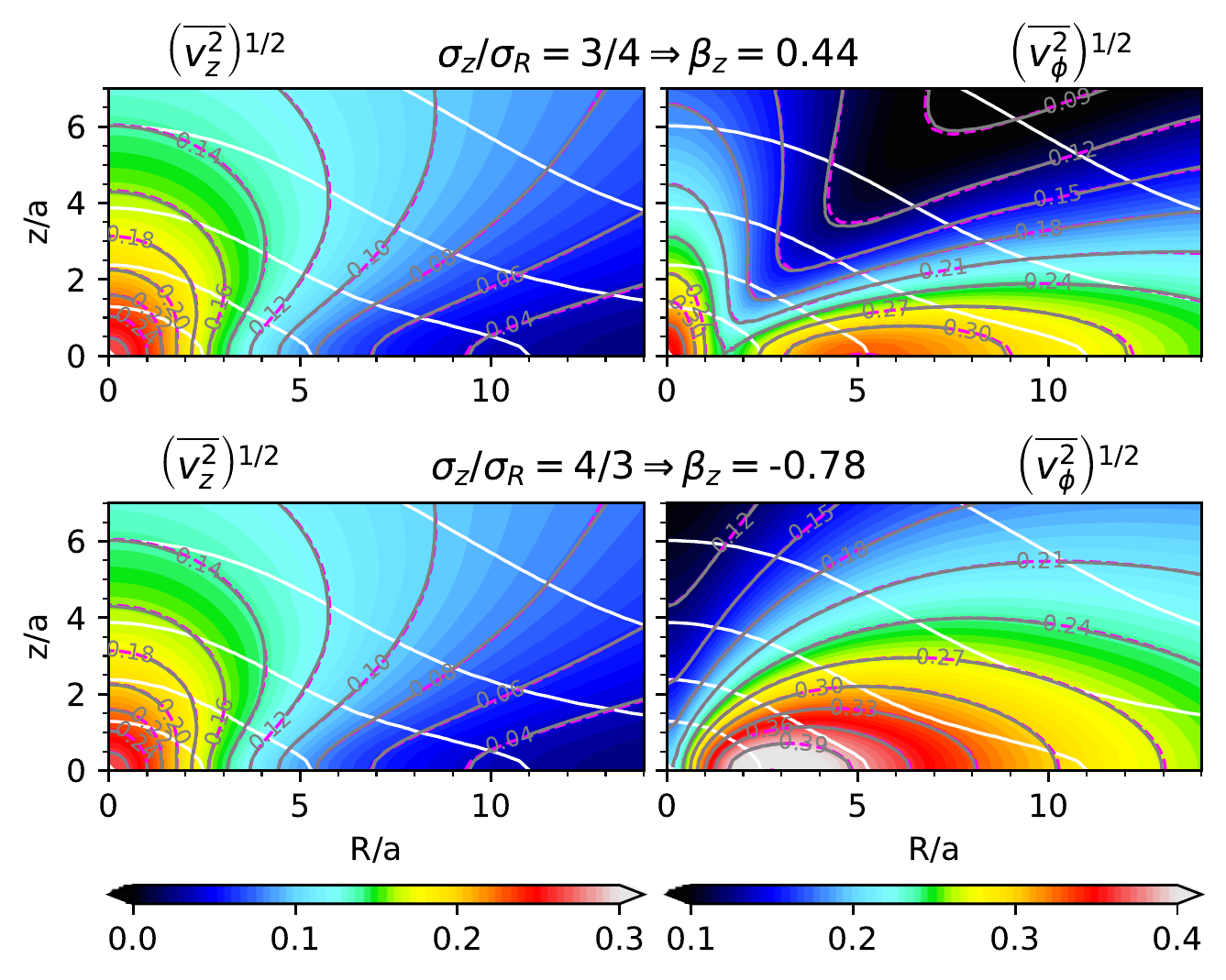}
	\caption{Intrinsic moments of JAM$_{\rm cyl}$ for two different anisotropies, for the Satoh's model of \autoref{fig:mge_satoh}. The colours and the grey contours with labels are the JAM$_{\rm cyl}$ solutions while the magenta dashed contours are the solutions of \autoref{sec:satoh_cyl}. The unit of velocity is $\sqrt{G M/a}$. The white contours are the model isodensity, spaced by factors of 10 starting from the maximum value. The anisotropy is written in the titles. This figure can be directly compared to the JAM$_{\rm sph}$ solution shown in \autoref{fig:jam_satoh}. \label{fig:jam_satoh_cyl}}
\end{figure}

In \autoref{fig:jam_atlas3d} I compare the $\overline{v_{\rm los}}$ and $\overline{v^2_{\rm los}}$ computed from both JAM$_{\rm cyl}$ of \citet{Cappellari2008} and JAM$_{\rm sph}$ presented in this paper. For the comparison, I selected the set of galaxies for which the JAM$_{\rm cyl}$ self-consistent models provides an excellent fit to the real data presented in fig.~10 of \citet{Cappellari2016}. From this set, I extracted the subset with significantly non-zero anisotropy $\beta_z\geq0.1$. The MGE models for these galaxies are taken from \citet{Scott2013p21}, while the best fitting model parameters\footnote{The model parameters and the tables with the MGEs are available from the ATLAS$^{\rm 3D}$ website \url{http://purl.org/atlas3d}} are taken from \citet{Cappellari2013p15}. For both models, I adopt the same MGE, the same inclination and $M/L$. I additionally adopt $\sigma_\theta/\sigma_r=\sigma_z/\sigma_R$, and $\sigma_\phi/\sigma_r=\sigma_\phi/\sigma_R$. In this way, the two sets of models have the same oblate shape of the velocity ellipsoid in the galaxies equatorial planes, where, by symmetry $\sigma_\theta=\sigma_z$ and $\sigma_r=\sigma_R$, while the shape of the two velocity ellipsoids gradually differs away from the equatorial plane.

The result of the qualitative comparison of \autoref{fig:jam_atlas3d} is that the two solutions look relatively similar, with differences roughly at the level one can expect from measurement errors in the stellar kinematics. The similarity is perhaps not surprising, given that the anisotropy of real fast rotator galaxies tends to be quite small, with typical values as measured from Schwarzschild models around $\beta\sim0.2$ \citep{Cappellari2007,Thomas2009}, and of course, JAM$_{\rm cyl}$ and JAM$_{\rm sph}$ must coincide in the isotropic limit.

The comparison using the rather small measured anisotropy of real galaxies should not give the misleading impression that JAM$_{\rm cyl}$ and JAM$_{\rm sph}$ remain close for any anisotropy. This is not the case. JAM$_{\rm sph}$ is characterized by a relative insensitivity of the model predictions to anisotropy. Instead, JAM$_{\rm cyl}$ quickly develops a vertical elongation in $\overline{v^2_{\rm los}}$, along the symmetry axis, for large positive $\beta_z$. This dramatic difference in the model behaviour is illustrated in \autoref{fig:jam_ngc4660_beta03}, where I construct models for one of the galaxies in \autoref{fig:jam_atlas3d} while adopting for both models an anisotropy that is significantly larger than that inferred using JAM$_{\rm cyl}$. While JAM$_{\rm sph}$ remains qualitatively similar to the solution in \autoref{fig:jam_atlas3d}, JAM$_{\rm cyl}$ becomes radically different and would be strongly inconsistent with the original fit (and the kinematic data in fig.~10 of \citealt{Cappellari2016}). 

\autoref{fig:jam_satoh_cyl} shows the intrinsic moments\footnote{Note that the left panel now shows $\overline{v_z^2}$ instead of $\overline{v_r^2}$. The two quantities are only comparable on the symmetry $z$-axis.} of JAM$_{\rm cyl}$ for the same Satoh's model and the same anisotropies as shown in \autoref{fig:jam_satoh} for JAM$_{\rm sph}$. The cylindrical solution for $\overline{v_z^2}$ in \autoref{eq:jeans_sol_z} is obviously independent of $\beta_z$. Instead, the solution for $\overline{v_\phi^2}$ shows a strong vertical elongation for $\beta_z=0.44$, which is the cause of the similar elongation in the $\overline{v^2_{\rm los}}$ for the projected moments in \autoref{fig:jam_ngc4660_beta03}. This radially anisotropic $\overline{v_\phi^2}$ solution also shows a diagonal depression (black colour in \autoref{fig:jam_satoh_cyl}), which, in this example, I found starts developing unphysical negative $\overline{v^2_\phi}$ values for $\beta_z>0.51$. 

\subsection{Which JAM method should one use?}

The availability of two different axisymmetric JAM$_{\rm sph}$ and JAM$_{\rm cyl}$ model implementations with either spherical or cylindrical alignment raise the question about which method one should use when studying real galaxies. In some cases, like for the outer stellar halo of the Milky Way, the answer is clear, given that we can measure the alignment of the velocity ellipsoid directly. However, for external galaxies, I have found that in general the two solutions can give quite comparable fits to the observed kinematics and it may not be clear which one provides the most reliable results for a certain quantity of interest.

My practical recommendation is {\em not} to prefer one over the other one, but instead to use {\em both} extreme assumptions on the alignment of the velocity ellipsoid made by the JAM$_{\rm cyl}$ and JAM$_{\rm sph}$ methods to asses the sensitivity of the model results to the model assumptions. When the two methods provide consistent results, one can be confident of derived physical quantities, while where the two methods differ, one should treat the results with caution. The difference between the results inferred using either JAM$_{\rm sph}$ or JAM$_{\rm cyl}$, especially when applied to large statistical samples, can be used as an estimate of the expected modelling errors.

The first application of this approach of comparing JAM$_{\rm sph}$ or JAM$_{\rm cyl}$ was presented in \citet{Nitschai2020}, which uses JAM to model the Gaia DR2 stellar kinematics and infer the mass distribution of the Milky Way. In that work we found that the two JAM methods give nearly-indistinguishable total density profiles, providing strong confidence in the derived result. An application to the statistically significant  ATLAS$^{\rm 3D}$ sample \citep{Cappellari2011a} of early-type galaxies is presented in the next section.

\subsection{Applying JAM$_{\rm sph}$ and JAM$_{\rm cyl}$ to the ATLAS$^{\rm 3D}$ sample}

\begin{figure}
	\includegraphics[width=\columnwidth]{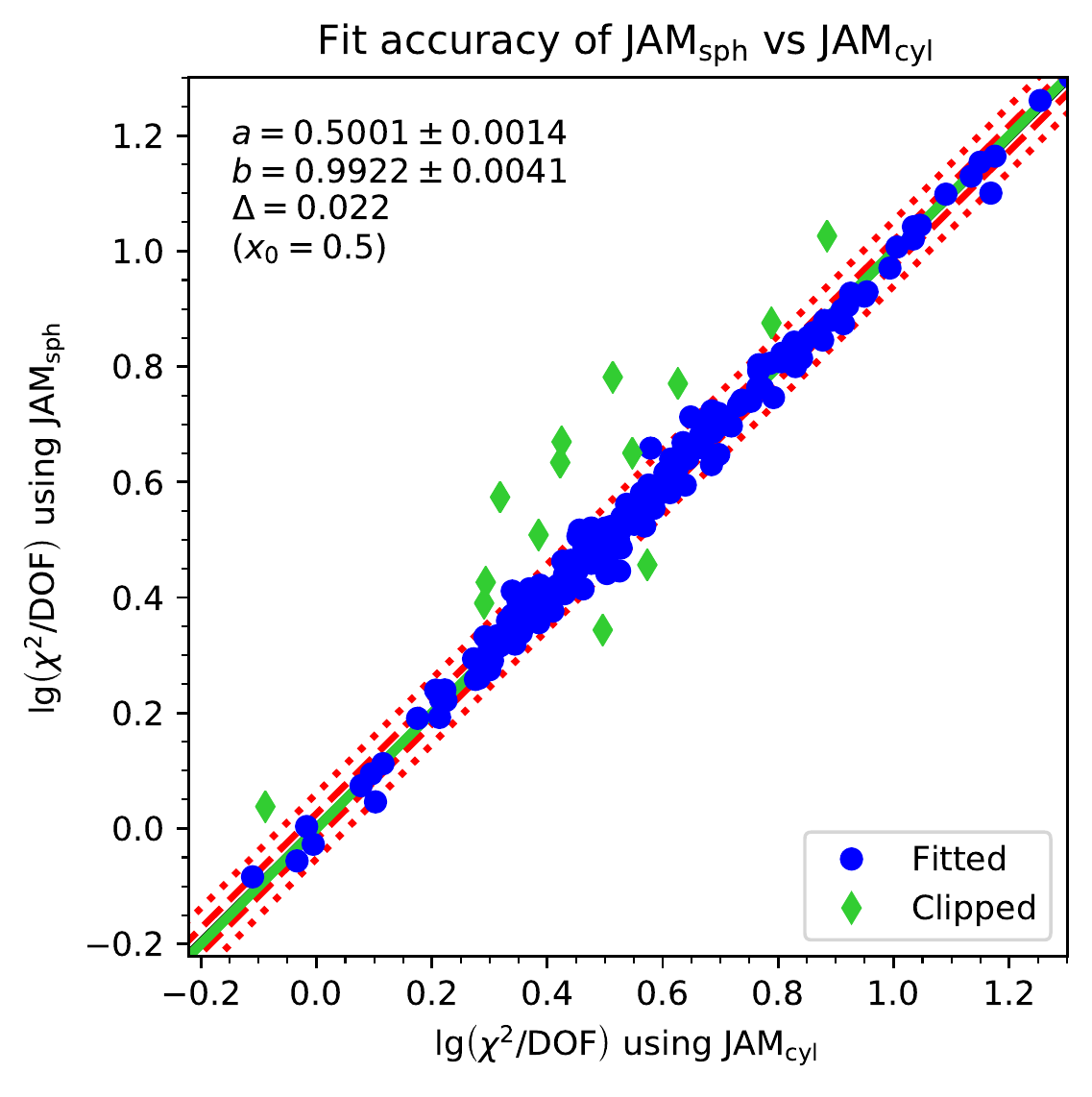}
	\caption{Comparison between the goodness-of-fit $\chi^2$ per degrees-of-freedom (DOF) obtained when fitting the $V_{\rm rms}$ kinematics of the ATLAS$^{\rm 3D}$ galaxies using either the JAM$_{\rm sph}$ or the JAM$_{\rm cyl}$ methods. The parameters of the best-fitting linear relation $y=a+b(x-x_0)$, the resulting $1\sigma$ uncertainties and the observed scatter $\Delta$ are printed in the top-left corner. The green line is the best-fitting relation while the dotted lines indicate the $1\sigma$ (68\% of values) and $2.6\sigma$ (99\% of values) scatter around the relation. The fit was performed with the robust \textsc{lts\_linefit} procedure by \citet{Cappellari2013p15} and the values clipped by the program are shown as green diamonds. \label{fig:fits_chi2_comparison}}
\end{figure}

As an illustration of how to use in practice the recommendation given in the previous section, here I applied both the JAM$_{\rm sph}$ and JAM$_{\rm cyl}$ methods to asses the robustness of the measurement of the total density slope for the whole ATLAS$^{\rm 3D}$ sample of early-type galaxies \citep{Cappellari2011a}, which were presented in \citet{Poci2017}. Even in this epoch, with the availability of the much larger MaNGA \citep{Bundy2015} and SAMI \citep{Bryant2015} integral-field spectroscopic (IFS) surveys, the ATLAS$^{\rm 3D}$ sample represents a useful and very well-studied benchmark due to its consistently high IFS data quality and higher spatial resolution. 

The modelling approach I used is the same as the model (D) in \citet{Cappellari2013p15} and I fitted the same $V_{\rm rms}$ kinematics from \citet{Cappellari2011a}. In brief, the models adopt a stellar component embedded in a spherical halo. The stellar components is parametrized by the MGE models\footnote{Both kinematics and MGEs are available from \url{http://purl.org/atlas3d}} from \citet{Scott2013p21}, with constant stellar mass-to-light ratio, while the halo density is described by a generalized NFW profile with free inner logarithmic slope \citep[gNFW,][]{Wyithe2001}. The fits of the JAM models to the kinematic data were performed with the \textsc{CapFit} constrained least-squares optimization program, which combines the Sequential Quadratic Programming and the Levenberg-Marquardt methods and is included in the \textsc{ppxf} Python package\footnote{I used v7.0 of the \textsc{ppxf} package from \url{https://pypi.org/project/ppxf/}} of \citet{Cappellari2017}. Data-model comparisons were already shown, for very similar models and the same data, in fig.~1 of \citet{Cappellari2013p15}, and the present fits are nearly indistinguishable from those.

\autoref{fig:fits_chi2_comparison} compares the goodness of fit $\chi^2/{\rm DOF}$ per degrees-of-freedom for both JAM$_{\rm sph}$ and JAM$_{\rm cyl}$. I perform a linear fit to the two quantities with the robust \textsc{lts\_linefit} procedure\footnote{I used v5.0 of the \textsc{LtsFit} package from \url{https://pypi.org/project/ltsfit/}} by \citet{Cappellari2013p15}, which combines the Least Trimmed Squares robust technique of \citet{Rousseeuw2006} into a least-squares fitting algorithm which allows for errors in both variables and intrinsic scatter. I find that the quality of the fits with the two methods is, on average, statistically indistinguishable, except for some outliers. 

After obtaining the best fits, I computed the resulting total-density average logarithmic slope $\gamma_{\rm tot}=\Delta\log\rho_{\rm tot}(r)/\Delta\log r$. I computed the spherically-averaged density $\rho_{\rm tot}(r)$ from the dark$+$luminous MGEs using the procedure \textsc{mge\_radial\_density} included in the JAM package, which implements the footnote~11 of \citet{Cappellari2015dm}. I considered a radial interval from 2 arcsec, which is a bit larger than the typical resolution of the kinematics, to the largest radius included in each IFS kinematics.

\begin{figure}
	\includegraphics[width=\columnwidth]{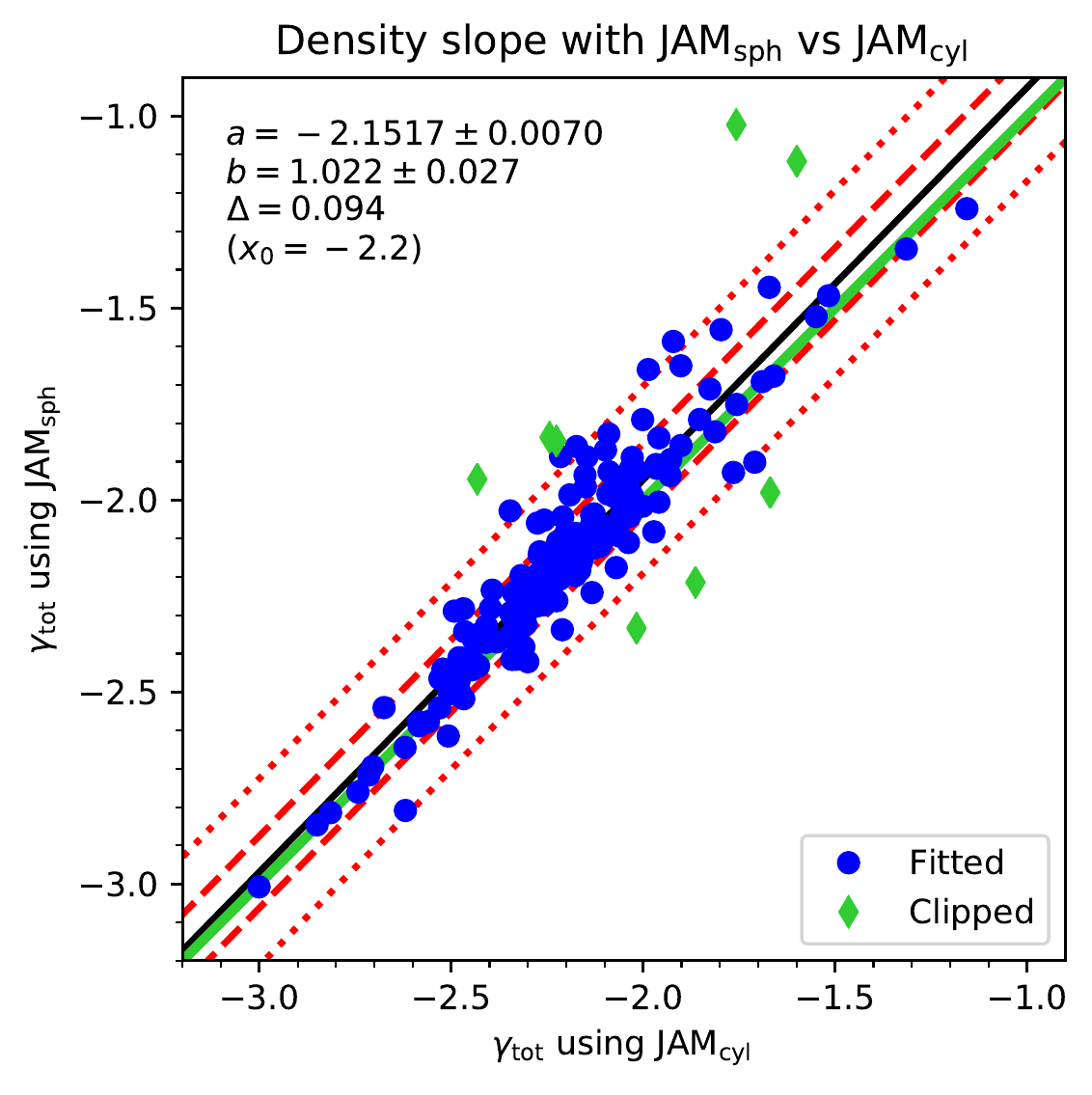}
	\caption{Same as in \autoref{fig:fits_chi2_comparison}, for the best-fitting total slopes $\gamma_{\rm tot}$ inferred using either the JAM$_{\rm sph}$ or the JAM$_{\rm cyl}$ methods to fit the ATLAS$^{\rm 3D}$ integral-field stellar kinematics. \label{fig:total_slope_comparison}}
\end{figure}

The resulting average density slope $\gamma_{\rm tot}$ inferred from the best-fitting models is shown in \autoref{fig:total_slope_comparison}. This too shows no systematic differences between the two JAM methods, except for some outliers. In the figure I only included galaxies with kinematic quality flag ${\rm qual}>0$ in table~1 of \citet{Cappellari2013p15}. The $\gamma_{\rm tot}$ derived with the two methods has an observed scatter $\Delta=0.094$. Assuming the uncertainties are the same for the two methods, this scatter implies a $1\sigma$ uncertainty of $\epsilon_\gamma=\Delta/\sqrt{2}\approx0.07$ in each slope determination. This value is close to the estimate $\epsilon_\gamma\approx0.09$ obtained by \citet{Poci2017}, confirming the validity of the approach.

The observed near insensitivity of the total slope inferred using either JAM$_{\rm sph}$ and JAM$_{\rm cyl}$ on real galaxies, namely its insensitivity to the assumed orientation of the velocity ellipsoid, appear to explain the accuracy of the total slopes previously reported for JAM$_{\rm cyl}$ (\autoref{sec:motivation}). 

\section{Conclusions}

I presented a general anisotropic solution for the axisymmetric Jeans equations of stellar hydrodynamics under the assumption of a velocity ellipsoid that is aligned with the spherical polar coordinate system. The solution requires a triple numerical quadrature with improper integrals for general gravitational potentials. I described an efficient and robust numerical method for its computation. The resulting algorithm is just one order of magnitude slower than my previously derived cylindrically-aligned solution, which only required a single quadrature. For reference, the computation of all components of the second velocity moment tensor and the mean velocities in \autoref{fig:jam_satoh_proj}, with my current Python implementation of the algorithm, took 7 s on a 2 GHz CPU.

I derived analytic equations for testing both the spherically-aligned and cylindrically-aligned anisotropic Jeans solutions and used them to verify the accuracy of both the formalism and the numerical implementations of the algorithms.

I described the general procedure and a method for the efficient numerical computation of the sky projection of all six components of the symmetric second velocity moment tensor and the three mean velocity components. I gave examples illustrating the qualitative trends in galaxy observables as a function of anisotropy.

I compared the spherically-aligned JAM$_{\rm sph}$ and cylindrically-aligned JAM$_{\rm sph}$ Jeans solutions using parameters describing the kinematics of real galaxies and found that for these cases the two methods produce rather similar observables, for the range of observed anisotropies, but can differ dramatically at larger anisotropy.

This JAM method has already been applied to model the Gaia DR2 data, where we found it describes the observations remarkably well with minimal freedom and good accuracy \citep{Nitschai2020}. 
Here, I also used both JAM$_{\rm sph}$ and JAM$_{\rm cyl}$ to model the ATLAS$^{\rm 3D}$ sample of early-type galaxies with high-quality integral-field stellar kinematics. I found that the inferred total-density slopes are nearly insensitive to the adopted orientation of the velocity ellipsoid and this appears to explain the previously-reported accuracy of JAM$_{\rm cyl}$ in recovering density profiles of real and simulated galaxies.

\input{cappellari2020_jam_spherically_aligned.bbl}

\appendix

\section{Cylindrically-aligned Axisymmetric Jeans solution}

This Appendix summarizes formulas from Section~3.1 of \citet{Cappellari2008}, for the LOS components, and from \citet{Cappellari2012jam}, for the proper motion components. All these expressions are also implemented in the publicly-available JAM software package.

\subsection{General solution}
\label{sec:jeans_solution}

Analogously to the procedure in \autoref{sec:jeans_sph}, one starts from the general axisymmetric Jeans equations in cylindrical coordinates and makes the following two assumptions: (i) the velocity ellipsoid is aligned with the cylindrical coordinate system $(R,\phi,z)$ and (ii) the anisotropy (of each MGE Gaussian) is constant and quantified by $\overline{v_R^2}=b\,\overline{v_z^2}$ (this implies $b=1/[1-\beta_z]$). In this case the Jeans equations reduce to
\begin{align}
\frac{b\,\nu\overline{v_z^2}-\nu\overline{v_\phi^2}}{R}
+ \frac{\partial(b\,\nu\overline{v_z^2})}{\partial R}
& = -\nu\frac{\partial\Phi}{\partial R}
\label{eq:jeans_beta_R}\\
\frac{\partial(\nu\overline{v_z^2})}{\partial z}
& = -\nu\frac{\partial\Phi}{\partial z},
\label{eq:jeans_beta_z}
\end{align}
which corresponds to the semi-isotropic case (two-integral) when $b=1$. With the boundary condition $\nu\overline{v_z^2}=0$ as $z\rightarrow\infty$
the solution reads
\begin{align}
\nu\overline{v_z^2}(R,z)
& =  \int_z^\infty \nu\frac{\partial\Phi}{\partial z}\dd z
\label{eq:jeans_sol_z}\\
\nu\overline{v_\phi^2}(R,z) & = 
b\left[
R \frac{\partial(\nu\overline{v_z^2})}{\partial R}
+ \nu\overline{v_z^2} \right]
+ R \nu\frac{\partial\Phi}{\partial R}
\label{eq:jeans_sol_R}.
\end{align}
A general caveat regarding the Jeans equations is that the existence of a solution does not guarantee the existence of a corresponding physical positive DF. As an example, the $\overline{v_\phi^2}$ can become non-physically negative for large $\beta_z$ as mentioned in \autoref{sec:sph_vs_cyl}

\subsection{MGE intrinsic quantities}
\label{sec:mge_solution}

In \citet{Cappellari2008} I applied the MGE formalism to the solution of the axisymmetric anisotropic Jeans equations of \autoref{sec:jeans_solution}. The resulting expressions for the intrinsic moments of each MGE Gaussian are
\begin{align}
[\nu\overline{v_R^2}]_k =\, & b_k[\nu\overline{v_z^2}]_k \label{eq:sigma_R}\\
[\nu\overline{v_{\phi}^2}]_k =\, & 4\pi G \int_0^1 \sum_{j=1}^M \frac{\nu_k q_j \rho_{0j} \left(\mathcal{D}\,R^2 + b_k \sigma_k^2 q_k^2\right) \mathcal{H}_j(u)\, u^2}{\mathcal{C}}
\dd u \label{eq:v2_phi_cyl}\\
[\nu\overline{v_z^2}]_k =\, & 4\pi G \int_0^1 \sum_{j=1}^M
\frac{\sigma_k^2 q_k^2 \nu_k q_j \rho_{0j} \mathcal{H}_j(u)\, u^2}
{\mathcal{C}}\, \dd u, \label{eq:sigma_z}
\end{align}
where 
\begin{align}
&\mathcal{C}=1-\left(1-q_j^2-\frac{\sigma_k^2\, q_k^2}{\sigma_j^2}\right)u^2,
\quad \mathcal{D}= \mathcal{C} - \frac{b_k\, q_k^2 (\sigma_j^2 + \sigma_k^2 u^2)}{\sigma_j^2}\\
&\mathcal{H}_j(u) = \frac{{\exp \left\{ - \frac{{u^2 }}
		{{2\sigma _j^2 }}\left[ {R^2  + \frac{{z^2 }}{{1 - (1 - q_j^2 )u^2 }}}
		\right] \right\}}}{{\sqrt {1 - (1 - q_j^2 )u^2 } }}.
\end{align}
Like before, the index $k$ refers to the parameters, or the anisotropy, of the Gaussians describing the galaxy's luminosity density of \autoref{eq:dens}, while the index $j$ refers to the parameters of the Gaussian describing the total mass of \autoref{eq:mass}, from which the potential is obtained.  These formulas generalized to anisotropic (three-integral) models what was done in the semi-isotropic (two-integral) self-consistent case ($b_k=1$ and $\rho_{0j}=\Upsilon \nu_{0k}$) by \citet{Emsellem1994}.

\subsection{MGE projected quantities}
\label{sec:jam_cyl_proj}

In \citet{Cappellari2008} I derived the cylindrically-aligned projected second velocity moments. I stated in note 5 that all these components can be written via single quadratures without the need for special functions, and I provided a reference software implementation, called the Jeans Anisotropic Modelling (\textsc{JAM}) method\footnote{Available from \url{https://pypi.org/project/jampy/}}. However, I only gave the line-of-sight component $\overline{v_{z'}^2}\equiv\overline{v_{\rm los}^2}$ in eq.~(28) of that paper. For completeness, I later provided all six components of the symmetric projected second velocity moment tensor in an addendum \citep{Cappellari2012jam}. The resulting formulas are reproduced in this Appendix. I updated them to conform to the new definition of the relation between galaxy's and observer's coordinates adopted in \autoref{eq:matrix_los} of this paper. Any of the six components of the symmetric projected second velocity moment tensor can be written as \citep{Cappellari2012jam}
\begin{align}\label{eq:second_moment}
\Sigma\,\overline{v_\alpha v_\beta}(x',y') =\, & 4\pi^{3/2} G \int_0^1 \sum_{k=1}^N \sum_{j=1}^M\, \nu_{0k}\, q_j\, \rho_{0j}\, u^2\, \mathcal{F}_{\alpha\beta} \nonumber \\
& \times\frac{
	\exp\left\{-\mathcal{A}\left[x'^2+y'^2 (\mathcal{A}+\mathcal{B})/\mathcal{E}\right]\right\}
}{\mathcal{C} \sqrt{\mathcal{E} \left[1-(1-q_j^2)u^2\right]}}\dd u, 
\end{align}
where $\alpha$ and $\beta$ stand for any of the three projected coordinates $x'$, $y'$ and $z'$, and I defined
\begin{align}
&\mathcal{A}=\frac{1}{2}\left(\frac{u^2}{\sigma_j^2} + \frac{1}{\sigma_k^2}\right),
\qquad\mathcal{E}=\mathcal{A}+\mathcal{B}\cos^2 i\\
&\mathcal{B}=\frac{1}{2}\left\{\frac{1-q_k^2}{\sigma_k^2 q_k^2}
+\frac{(1-q_j^2)u^4}{\sigma_j^2\left[1-(1-q_j^2)u^2\right]}\right\}.
\end{align}
The expressions for the projection factors $\mathcal{F}_{\alpha\beta}$ are
\begin{subequations}
	\begin{align}
	\mathcal{F}_{x'x'}&=b_k \sigma_k^2 q_k^2 + \mathcal{D}\, \left\{\left[y'\cos i\, (\mathcal{A}+\mathcal{B})/\mathcal{E}\right]^2+\sin^2 i/(2\mathcal{E})\right\}\\
	\mathcal{F}_{y'y'}&=\sigma_k^2 q_k^2 \left(\sin^2 i + b_k \cos^2 i\right) + \mathcal{D}\, x'^2\cos^2 i\\
	\mathcal{F}_{z'z'}&=\sigma_k^2 q_k^2 \left(\cos^2 i + b_k \sin^2 i\right) + \mathcal{D}\, x'^2\sin^2 i\\
	\mathcal{F}_{x'y'}&=-\mathcal{D}\, x'y'\cos^2 i\, (\mathcal{A}+\mathcal{B})/\mathcal{E},\\
	\mathcal{F}_{x'z'}&=\mathcal{F}_{x'y'}\tan i=-\mathcal{D}\, x'y'\sin i\,\cos i\, (\mathcal{A}+\mathcal{B})/\mathcal{E},\\
	\mathcal{F}_{y'z'}&= \sin i\, \cos i\, \left[\mathcal{D}\, x'^2 - \sigma_k^2 q_k^2\, (1-b_k)\right].
	\end{align}
\end{subequations}
The expressions for $\overline{v^2_{x'}}$ and $\overline{v^2_{y'}}$ where also given in \citet{DSouza2013}. And the whole derivation was summarized in detail by \citet{Watkins2013}. 

The procedure to compute the projected first velocity moments $\overline{v_{x'}}$, $\overline{v_{y'}}$ and $\overline{v_{z'}}\equiv\overline{v_{\rm los}}$ is identical in this cylindrically-oriented case to the spherically-oriented one. In both cases, the only non-zero component of the mean velocity is the $\overline{v_{\phi}}$ component. No analytic LOS integral seems possible in this case and the LOS integration is performed with an extra numerical quadrature, by (i) first computing the mean velocity $\overline{v_{\phi}}$ using \autoref{eq:vphi}, for an adopted splitting of $\overline{v_\phi^2}$, (ii) then projecting the $\overline{v_{\phi}}$ along the desired component using \autoref{eq:los_v} and (iii) finally integrating the projected mean velocity along the LOS with \autoref{eq:los_integ_v}. The same numerical implementation approach described in \autoref{sec:implementation}, to exploit the axisymmetry of the solution, and the same TANH variable transformation, can be used also here to speed up the numerical calculation.

\section{Spherical Jeans solution}

This Appendix summarizes formulas from Section~3.2 of \cite{Cappellari2008}, for the LOS components, and from \citet{Cappellari2015jam}, for the components of the proper motion. All these expressions are also implemented in the publicly-available JAM software package.

\subsection{General solution}

Starting from \autoref{eq:jeans_sph_r} and assuming spherical symmetry one can obtain the Jeans equation as (\citealt{Binney1980}; equation~[4-54] of BT)
\begin{equation}
\frac{\dd(\nu\overline{v_r^2})}{\dd r}+\frac{2\beta\, \nu\overline{v_r^2}}{r}=-\nu\frac{\dd \Phi}{\dd r},
\end{equation}
where $\overline{v_{\theta}^2}=\overline{v_{\phi}^2}$ for symmetry and I defined $\beta=1-\overline{v_{\theta}^2}/\overline{v_r^2}$.
The solution of this linear first-order differential equation with constant anisotropy $\beta$ and the boundary condition $\nu\overline{v_r^2}=0$ as $r\rightarrow\infty$ is \citep[e.g.][]{Binney1980,Tonry1983,vanDerMarel1994m87}
\begin{align}\label{eq:spherical_jeans}
\nu\overline{v_r^2}(r) &= 
\int_r^{\infty} \left(\frac{r'}{r}\right)^{2\beta}
\nu(r')
\frac{\dd\Phi(r')}{\dd r'}\dd r'\nonumber\\
& = G\, \int _r^{\infty } \left(\frac{r'}{r}\right)^{2\beta} \frac{\nu(r')M(r')}{r'^2}\dd r',
\end{align}
considering that $\dd\Phi/\dd r=G M/r^2$. 

\subsection{MGE intrinsic quantities}
\label{sec:mge_spherical}

To evaluate the solution of \autoref{eq:spherical_jeans}, one needs to make a choice for the tracer and mass distributions. In \citet{Cappellari2008} I adopted for both the spherical MGE parametrization. In this case the surface brightness $\Sigma_k$, the luminosity density $\nu_k$ and the total density $\rho_j$ for each individual Gaussian are given by \citep{Bendinelli1991}
\begin{align}\label{eq:surf_sph}
\Sigma_k(R) &=
\frac{L_k}{2\pi\sigma^2_k}
\exp\left(-\frac{R^2}{2\sigma^2_k}\right),\\
\nu_k(r) &=
\frac{L_k}{\left(\sqrt{2\pi}\, \sigma_k\right)^3}
\exp\left(-\frac{r^2}{2\sigma_k^2}\right),\label{eq:dens_sph}\\
\rho_j(r) &=
\frac{M_j}{\left(\sqrt{2\pi}\, \sigma_j\right)^3}
\exp\left(-\frac{r^2}{2\sigma_j^2}\right).\label{eq:mass_sph}
\end{align}
The mass of a Gaussian contained within the spherical radius $r$ is given by equation~(49) of \citet{Cappellari2008}
\begin{equation}\label{eq:massr}
M_j(r)=M_j\times \left[{\rm erf}\left(\frac{r}{\sqrt{2}\, \sigma_j}\right)
-\frac{r\,\sqrt{2/\pi}}{\sigma_j}
\exp\left(-\frac{r^2}{2 \sigma_j^2}\right)
\right],
\end{equation}
with ${\rm erf}(x)$ the error function (\href{https://dlmf.nist.gov/7.2.E1}{equation~7.2.1} of \citealt{Olver2010nist}). Computing \autoref{eq:spherical_jeans} requires a single numerical quadrature.

\subsection{MGE projected quantities}

Following the same steps and definitions as for the line-of-sight velocity component \citep[sec.~3.2.1]{Cappellari2008} one can write the projection expressions for all three components of the velocity second moments, including the proper motions as follows
\begin{eqnarray}
\Sigma \overline{v_{\alpha}^2}(R)=
2G\int_R^{\infty }
\left[\frac{r^{1-2 \beta } \mathcal{Q}_\alpha(r)}{\sqrt{r^2-R^2}}
\int _r^{\infty }\frac{\nu(u)M(u)}{u^{2-2\beta}}\dd u\right] \, \dd r,
\end{eqnarray}
where (i) $\alpha={\rm los}$ for the line-of-sight velocity (ii) $\alpha={\rm pmr}$ for the  radial proper motion, measured from the projected centre of the system, and (iii) $\alpha={\rm pmt}$ for the tangential proper motion. The projection factors $\mathcal{Q}_\alpha$ are \citep{Leonard1989,Strigari2007,vanderMarel2010}
\begin{subequations}
	\begin{align}
	\mathcal{Q}_{\rm los}(r) &=1-\beta\, (R/r)^2\\
	\mathcal{Q}_{\rm pmr}(r) &=1-\beta+\beta\, (R/r)^2\\
	\mathcal{Q}_{\rm pmt}(r) &=1-\beta.
	\end{align}
\end{subequations}

Integrating by parts one of the two integrals vanishes and all three projected second moments can still be written as in equation~(42) of \citet{Cappellari2008}
\begin{equation}\label{eq:sph_jeans}
\Sigma \overline{v_\alpha^2}(R) = G\int_R^{\infty}
\mathcal{F}_\alpha \left(\frac{R^2}{r^2}\right)\nu(r)\,M(r)\, \dd r.
\end{equation}
When using the MGE parametrization, the evaluation of this expression requires a single numerical quadrature and some special functions. The expressions for all three components of $\mathcal{F}_\alpha$ are
\begin{subequations}
	\begin{align}\label{eq:f_los}
	\mathcal{F}_{\rm los}(w) =\, & \mathcal{A} - \mathcal{B}\\
	\mathcal{F}_{\rm pmr}(w) =\, & (1 - \beta)\,\mathcal{A} + \mathcal{B}\\
	\mathcal{F}_{\rm pmt}(w) =\, & (1 - \beta)\,\mathcal{A}
	\end{align}
\end{subequations}
with
\begin{subequations}\label{eq:ab_proper_motion}
	\begin{align}
	&\mathcal{A} = \frac{w^{1-\beta }}{R}
	\left[
	\frac{\sqrt{\pi }\, \Gamma\left(\beta-\frac{1}{2} \right)}{\Gamma(\beta )}
	- B_w \left(\beta-\frac{1}{2} ,\frac{1}{2}\right) 
	\right]\\
	&\mathcal{B} = \frac{w^{1-\beta }}{R}
	\left[
	\frac{\sqrt{\pi }\, \Gamma\left(\beta+\frac{1}{2} \right)}{\Gamma(\beta )}
	- \beta\, B_w \left(\beta+\frac{1}{2} ,\frac{1}{2}\right) 
	\right],
	\end{align}
\end{subequations}
where $\Gamma$ is the Gamma function (\href{https://dlmf.nist.gov/5.2.E1}{equation~5.2.1} of \citealt{Olver2010nist}) and $B_w$ is the incomplete Beta function (\href{https://dlmf.nist.gov/8.17.E1}{equation~8.17.1} of \citealt{Olver2010nist}), for which efficient routines exist in virtually any language. 
Specific expressions can be obtained for $\beta=\pm1/2$, where the $B_w$ function is divergent, but these expressions are not useful in real applications as it is sufficient to perturb $\beta$ by a negligible amount to avoid the singularity.
The expression for the line-of-sight component $\mathcal{F}_{\rm los}$ was given by \citet{Mamon2005} and I unknowingly re-derived it in equation~(43) of \citet{Cappellari2008}, while the formulas for the two proper motion components were given in \citet{Cappellari2015jam}.

The projected second velocity moments for the whole MGE model, summed over all the $N$ luminous and $M$ massive Gaussians, for any of the three velocity second moment components, are still given by equation~(50) of \citet{Cappellari2008}
\begin{equation}\label{eq:sph_jeans_mge}
\Sigma \overline{v_\alpha^2}(R) = G \int_R^{\infty}
\sum_{k=1}^N \mathcal{F}_{\alpha,k} \left(\frac{R^2}{r^2}\right)\nu_k(r)
\left[M_\bullet + \sum_{j=1}^M M_j(r)\right]
\dd r,
\end{equation}
where $\nu_k(r)$ is given by \autoref{eq:dens_sph}, $M_j(r)$ is given by \autoref{eq:massr}, and $\mathcal{F}_{\alpha,k}$  is obtained by replacing the $\beta$ parameter in \autoref{eq:ab_proper_motion} with the anisotropy $\beta_k$ of each luminous Gaussian component of the MGE.

\bsp    
\label{lastpage}
\end{document}